\documentclass[aps,prb,preprint]{revtex4-2}

\usepackage[utf8]{inputenc}

\usepackage{amsmath,amsthm,amssymb,amsfonts,dsfont} 

\usepackage{siunitx,microtype,hyperref}

\usepackage{graphicx,booktabs,rotating,colortbl,xcolor}

\usepackage{enumitem}

\usepackage[caption=false]{subfig}
 
\newcommand{\Prob}{\mathbb{P}}
\newcommand{\Ex}{\mathbb{E}}
 
\newcommand{\E}{\mathcal{E}}

\DeclareMathOperator*{\argmin}{argmin}

\DeclareMathOperator*{\Var}{Var}

\def\ch#1{{#1}}
\def\chd#1{}

\def\ds{deep silicon}
\def\ds{edge-on-irradiated silicon}

\begin{document}

\title{The effects of intra-detector Compton scatter on zero-frequency DQE for photon-counting CT using edge-on-irradiated silicon detectors}

\author{Fredrik Grönberg}
\email{gronberg@mi.physics.kth.se}
\affiliation{Physics of Medical Imaging, KTH Royal Institute of Technology, AlbaNova University Center, SE-106 91 Stockholm, Sweden}

\author{Zhye Yin}
\affiliation{GE Healthcare, Niskayuna, New York 12309, USA}

\author{Jonathan Maltz}
\affiliation{GE Healthcare, Waukesha, Wisconsin 53188, USA}

\author{Norbert J. Pelc}
\affiliation{Department of Radiology, Stanford University, Stanford, California, USA}

\author{Mats Persson}
\affiliation{\phantom{}Physics of Medical Imaging, KTH Royal Institute of Technology, AlbaNova University Center, SE-106 91 Stockholm, Sweden}

\begin{abstract}
\noindent
\textbf{Background:} Edge-on-irradiated silicon detectors are currently being investigated for use in full-body photon-counting CT applications. The low atomic number of silicon leads to a significant number of incident photons being Compton scattered in the detector, depositing a part of their energy and potentially being counted multiple times. Even though the physics of Compton scatter is well established, the effects of Compton interactions in the detector on image quality for an \ds\ detector have still not been thoroughly investigated.
\\
\\
\textbf{Purpose:} To investigate and explain effects of Compton scatter on zero-frequency DQE for photon-counting CT using \ds\ detectors. 
\\
\\
\textbf{Methods:} We extend an existing Monte Carlo model of an \ds\ detector, previously used to evaluate spatial-frequency-based performance, to develop projection and image domain performance metrics for pure density and pure spectral imaging tasks. We show that the lowest energy threshold of the detector can be used as an effective discriminator of primary counts and cross-talk caused by Compton scatter. We study the developed metrics as functions of the lowest threshold energy. We also compare the performance of a modeled detector with 8 optimized energy bins to a detector with an unlimited number of bins. 
\\
\\
\textbf{Results:} Density imaging performance as a function of the lowest threshold compared to an ideal photon-counting detector decreases monotonically at an approximate rate of 1.7 percentage units per keV, with a maximum of 0.68--0.69 for a lowest threshold of 0~keV and a value of 0.51 at 10~keV, in both projection and image domains. Spectral imaging performance compared to an ideal photon- counting detector has a plateau between 0 and 10~keV with values of 0.26--0.27 with 8 energy bins in both projection and image domain, and 0.28 with unlimited energy bins in the projection domain. Between 10~keV and 30~keV, the spectral imaging performance decreases with a rate of approximately 0.7 percentage units per keV.
\\
\\ 
\textbf{Conclusions:}
Compton interactions contribute significantly to the density imaging performance of \ds\ detectors. With the studied detector topology, the benefit of counting primary Compton interactions outweighs the penalty of multiple counting at all lower threshold energies. Compton interactions also contribute significantly to the spectral imaging performance for measured energies above 10~keV. 
\\
\\
\textbf{Key words:} photon-counting, x-ray detector
\end{abstract}

\maketitle

\section{Introduction}\label{sec:Introduction}

Photon-counting detectors (PCDs) are expected \ch{to yield the next major} advance in x-ray computed tomography (CT), with potential improvements in dose efficiency, spatial resolution, and spectral imaging compared to the current state-of-the-art. PCDs count single photons and measure their energy, compared to conventional energy-integrating detectors, which measure the total energy in the x-ray beam. \ch{Benefits of photon-counting are} improved contrast-to-noise ratio (CNR) and dose efficiency. The use of direct conversion semiconductors in PCDs eliminates the need for inter-pixel reflectors and therefore enables PCDs to have smaller pixels and higher spatial resolution, without reducing dose efficiency. The energy resolution \ch{can} enable better spectral imaging performance than what can be achieved with existing dual-energy CT technologies.

There are mainly two detector materials currently being investigated for use in full-body photon-counting CT applications\ch{:} silicon and CdTe/CZT. Both materials have non-ideal properties that affect performance. Silicon, since it has low atomic number, is relatively transparent to x-rays and thus requires a thicker sensor, on the order of 30--80~mm, to achieve a clinically acceptable dose efficiency. This can be achieved with a detector topology with sensors mounted \ch{``}edge-on\ch{''}, with their edge directed towards the x-ray source, and multiple channels along the depth of the sensors. \ch{The latter, along with the inherent speed of silicon detectors, allow} the detector to handle high count rates without reducing the pixel size to the \ch{point where} charge sharing becomes a dominating effect. The low atomic number of silicon also means that a significant number of incident photons will deposit energy in Compton interactions, scattering inside the detector and potentially being counted multiple times. Multiple counting of incident photons is detrimental to image quality, \ch{as has been previously established} \cite{michel2006fundamental}. With regard to spectral performance, Compton interactions provide less information than photoelectric interactions. However, due to the Compton edge (\ch{for a given incident energy, there is a maximum energy that can be deposited}), and prior information about the incident spectrum, some spectral information remain\ch{s within the} Compton interactions. There is further benefit that, unlike for distortion mechanisms such as charge sharing and K-fluorescence, Compton events are non-overlapping with the photopeak part of the deposited spectrum.

Even though the physics of Compton scatter is well established, the effects \ch{of Compton interactions} on image quality for a \ds\ detector \ch{have} still not been thoroughly investigated. In this paper, we build on previously published work on a Monte Carlo model of \ds\ sensors \cite{persson2020detective}. \ch{This model} has been used to evaluate frequency-based performance metrics, showing that silicon has the potential to be a competitive detector material for photon-counting CT. The purpose of this paper is to investigate the effect of Compton scatter in \ds\ detectors using simulation experiments \ch{by means of} projection and image domain detective quantum efficiency (DQE) metrics for both density and spectral imaging tasks. DQE is defined as the optimal $\text{CNR}^2$ for a particular contrast task relative to that of an ideal photon-counting detector.

Several previous works have described detailed simulation models for photon-counting x-ray detectors and used these to predict imaging performance for different detector designs in the projection domain \cite{taguchi2016spatio, taguchi2018spatio,rajbhandary2018effect} and the image domain.\cite{faby2016efficient} Several such previous studies are based on linear systems analysis and use DQE as a performance metric\cite{tanguay2015detective, persson2018framework,stierstorfer2018modeling, rajbhandary2020detective}. However, publications investigating the effect of Compton scatter in \ds\ detectors on DQE are few and limited to projection domain studies \cite{bornefalk2010photon,persson2020detective}. 

The remainder of this paper is structured as follows. In Section \ref{sec:Preliminaries} we provide a high-level description of the effects of Compton scatter in \ds\ detectors by describing the edge-on detector topology and the particular detector design studied in this work (\ref{sec:detector_topology}), showing and describing simulated response characteristics of this detector (\ref{sec:response_characteristics}), and deriving the impact of count multiplicity on a photon-counting detector without energy resolution (\ref{sec:count_multiplicity}. In Section \ref{sec:Detector_Simulation} we describe how the spatio-energetic point-spread function (PSF) and autocovariance function (ACF) of the detector were obtained from Monte Carlo simulations and in Sections \ref{sec:projection_evaluation} and \ref{sec:image_evaluation}, we describe how \ch{these} were used to obtain DQE metrics for density and spectral imaging tasks in the projection and image domain\ch{s}. Sections \ref{sec:Results}, \ref{sec:Discussion} and \ref{sec:Conclusion} present our results, discussion and conclusion, respectively. 

\section{Methods}\label{sec:Methods}

  \subsection{Preliminaries}\label{sec:Preliminaries}

    \subsubsection{Detector topology}\label{sec:detector_topology}

\begin{figure*}
      \centering
      \includegraphics[scale=0.45]{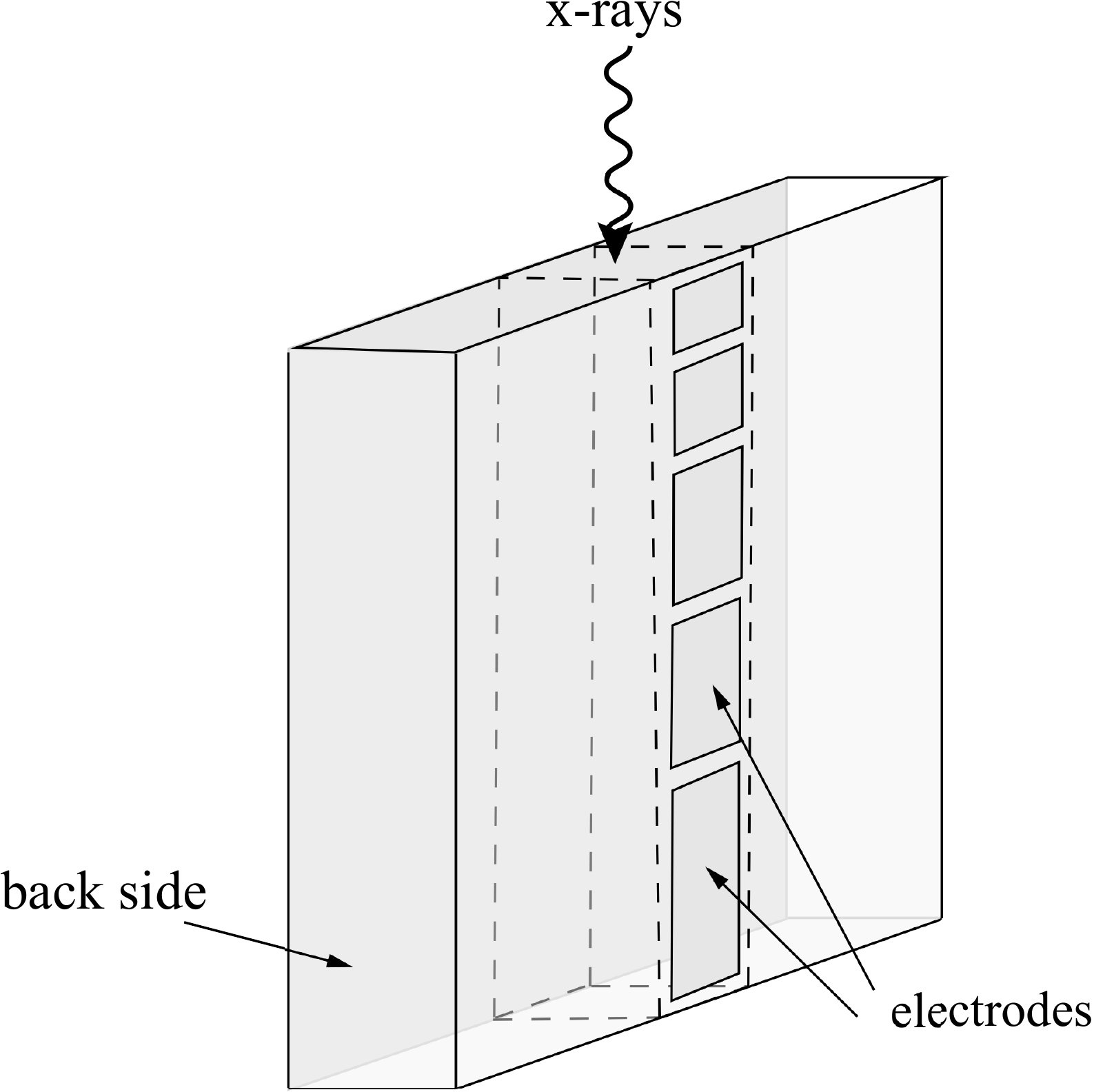}
      \caption{Drawing of a \ds\ detector module mounted in an edge-on geometry. By stacking a large number of these, a detector \ch{that covers a arbitrarily large  area}  can be assembled, similar to the detector simulated in this study. For simplicity, the detector simulated here does not have depth segmented electrodes. Image reproduced from \cite{danielsson2021photoncounting} with permission.}
      \label{fig:edgeon}
      \end{figure*}

      The \ds\ detector design studied here has been described previously \cite{persson2020detective} and is similar to detector prototypes that have been \ch{fabricated}, and evaluated in experimental studies. The fundamental building block of this detector is a silicon wafer covered by \ch{a} large electrode on one side, and by an array of strip-shaped electrodes on the other, forming a one-dimensional array of diodes serving as detector pixels (Fig. \ref{fig:edgeon}). To achieve sufficiently large detection efficiency, the detector is mounted in an edge-on configuration, so that the x-ray photons are attenuated through several centimeters of silicon. By stacking a large number of these wafers next to each other, a two-dimensional pixel array with a geometric detection efficiency close to unity can be achieved.

      Photons that undergo Compton scattering in the detector are, for the most part, deflected \ch{so as} to leave the original pixel of incidence, and often also the wafer of incidence. \ch{These deflected photons} are potentially reabsorbed in another pixel and thus cause image blur. This blurring, together with the effect of Compton scatter on spectral response, can be corrected \ch{via} deconvolution with the known detector response. Reabsorption can also lead to a photon being counted twice \ch{in cases where} both interactions deposit enough energy to be detected above the lowest energy threshold, and this degrades the CNR of the resulting image. For this reason, foils of a highly attenuating material, such as tungsten, are typically interspersed between the silicon wafers in order to absorb photons that have been scattered in the silicon\ch{.} \ch{This} reduces the amount of double counting \cite{bornefalk2010photon}. It is important to note that the possibility of double counting does not mean that counting Compton interactions is detrimental in general. Many Compton-interacting photons subsequently escape the sensor or are absorbed in the foils, meaning that the Compton part of the spectrum must be measured in order to capture these photons.

      The \ds\ detector design simulated in this paper has a pixel pitch of 500~$\mu$m in each of the $x$ and $y$ directions, and \ch{consists of} a stack of silicon wafers separated by a one-dimensional grid of 20~$\mu$m thick tungsten foils parallel to the transaxial plane, giving a fill factor of 0.96. The studied detector design had a measuring absorption length of 60~mm with a dead layer of 0.5~mm silicon at the entrance and exit plane of the wafer. These \ch{layers model the guard ring that is used to minimize} leakage current.

    \subsubsection{Response characteristics}\label{sec:response_characteristics}

\begin{figure*}
        \centering
        \begin{minipage}[b]{\linewidth}
        \subfloat[]{\includegraphics[width=0.39\linewidth]{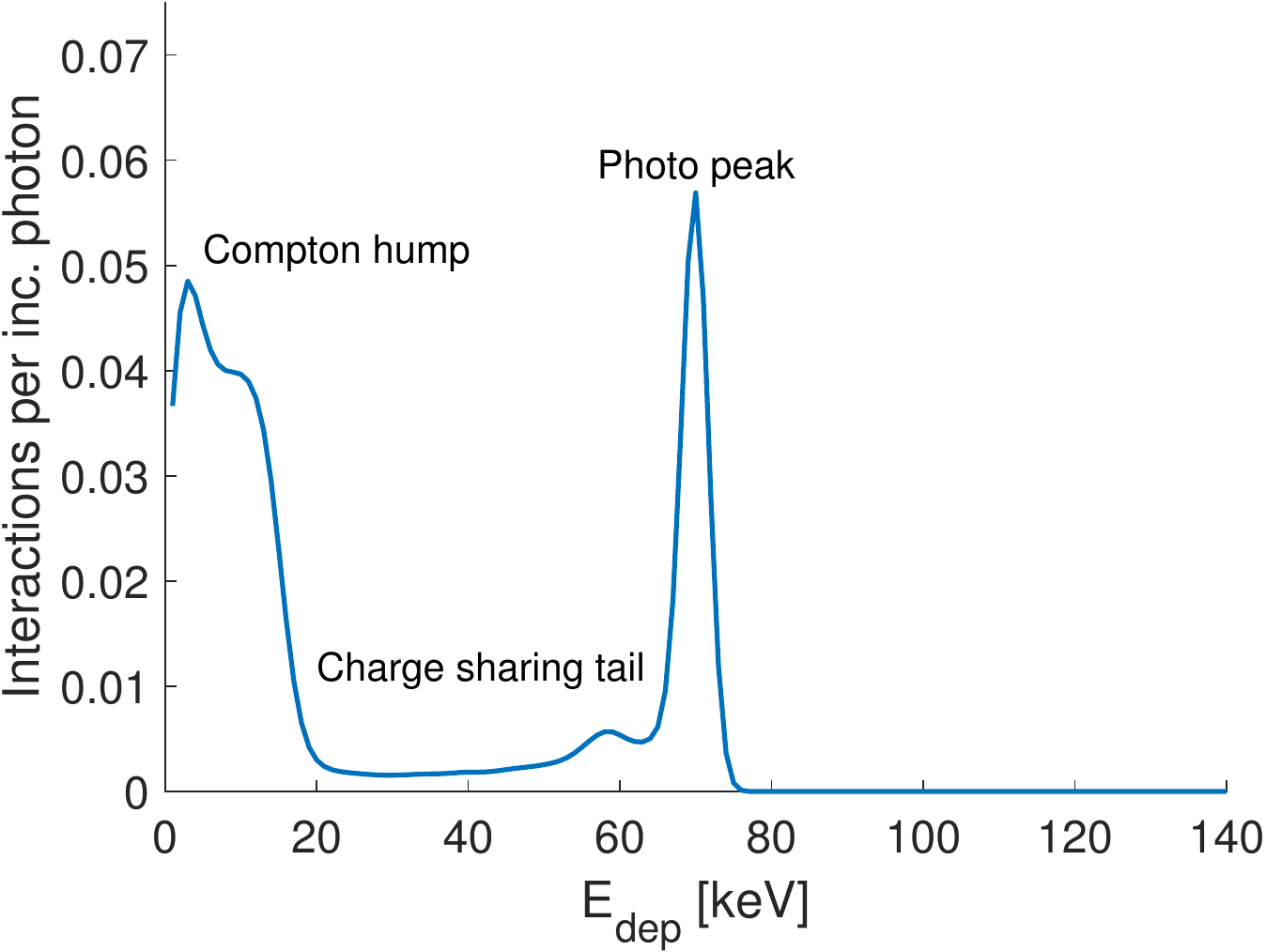}}
        \hfil
        \subfloat[]{\includegraphics[width=0.39\linewidth]{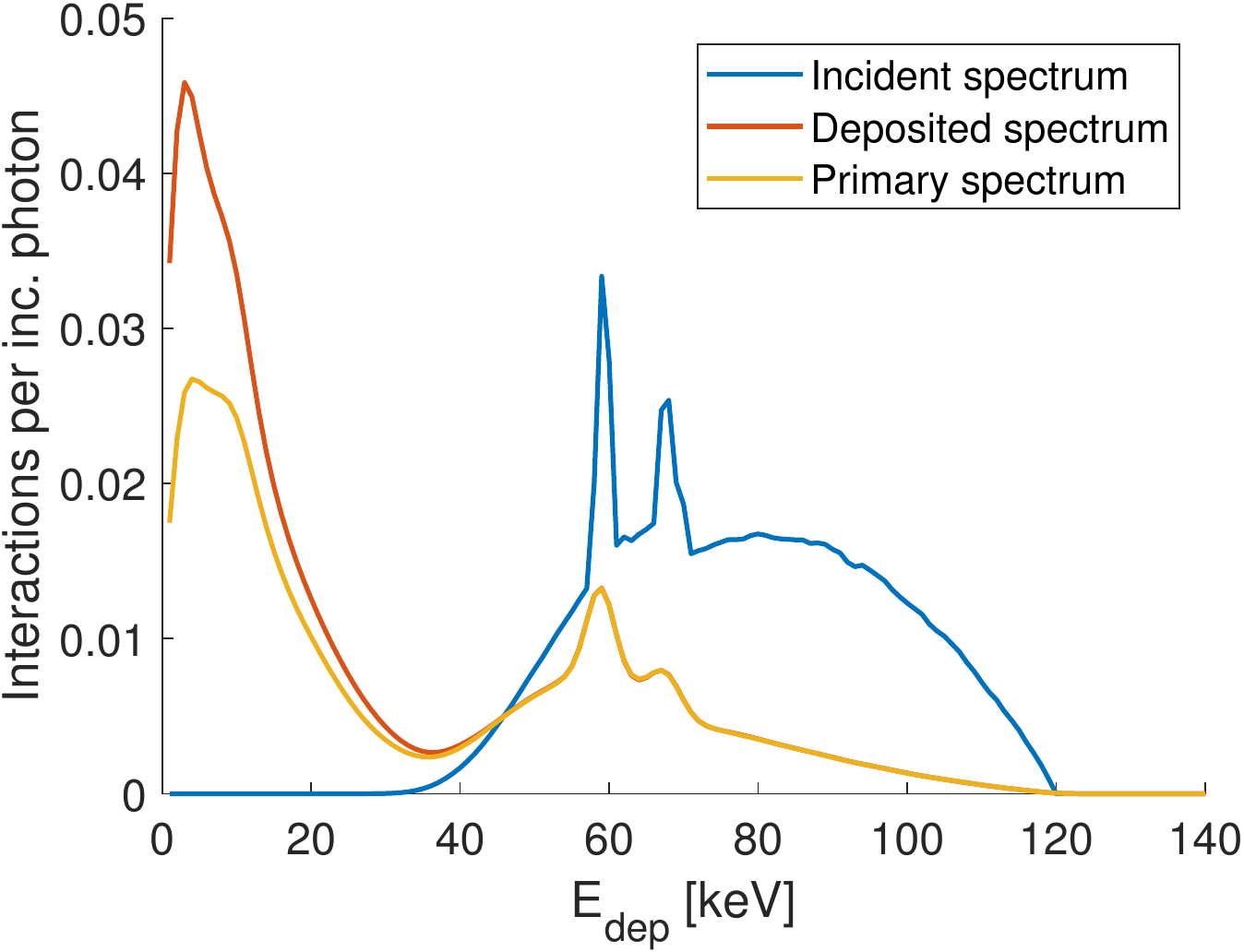}}
        \vfil
        \subfloat[]{\includegraphics[width=0.39\linewidth]{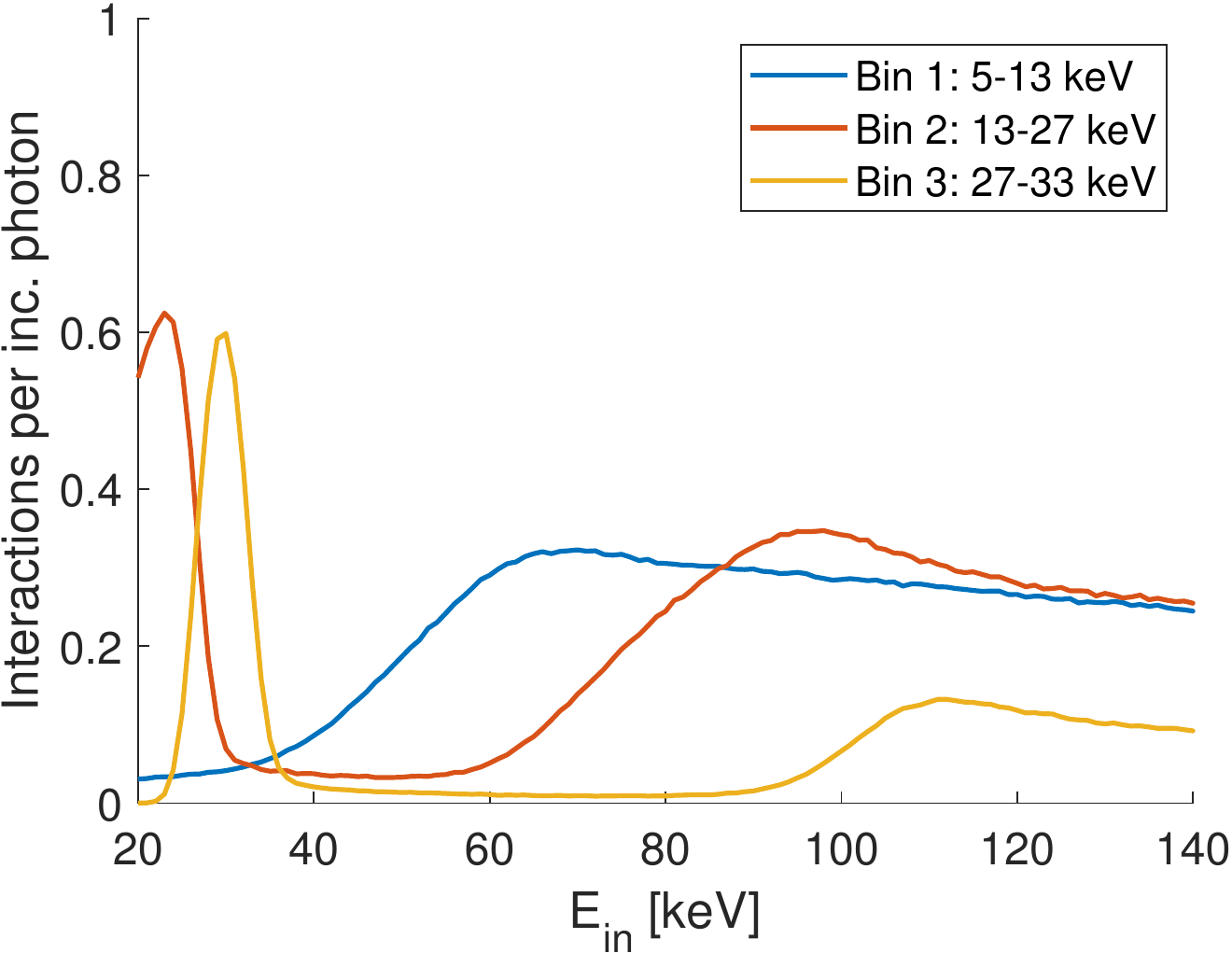}}
        \hfil
        \subfloat[]{\includegraphics[width=0.39\linewidth]{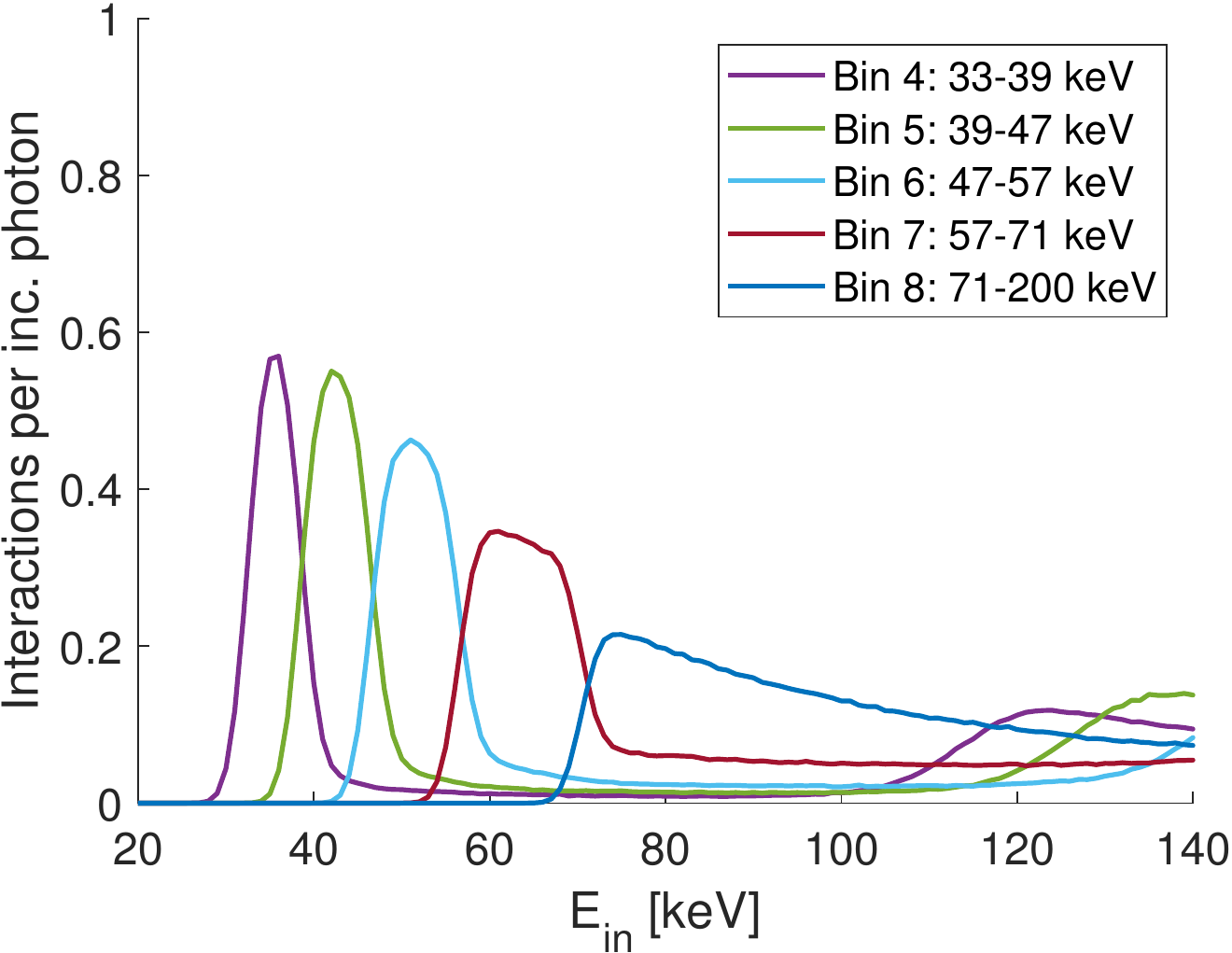}}
        \vfil
        \subfloat[]{\includegraphics[width=0.39\linewidth]{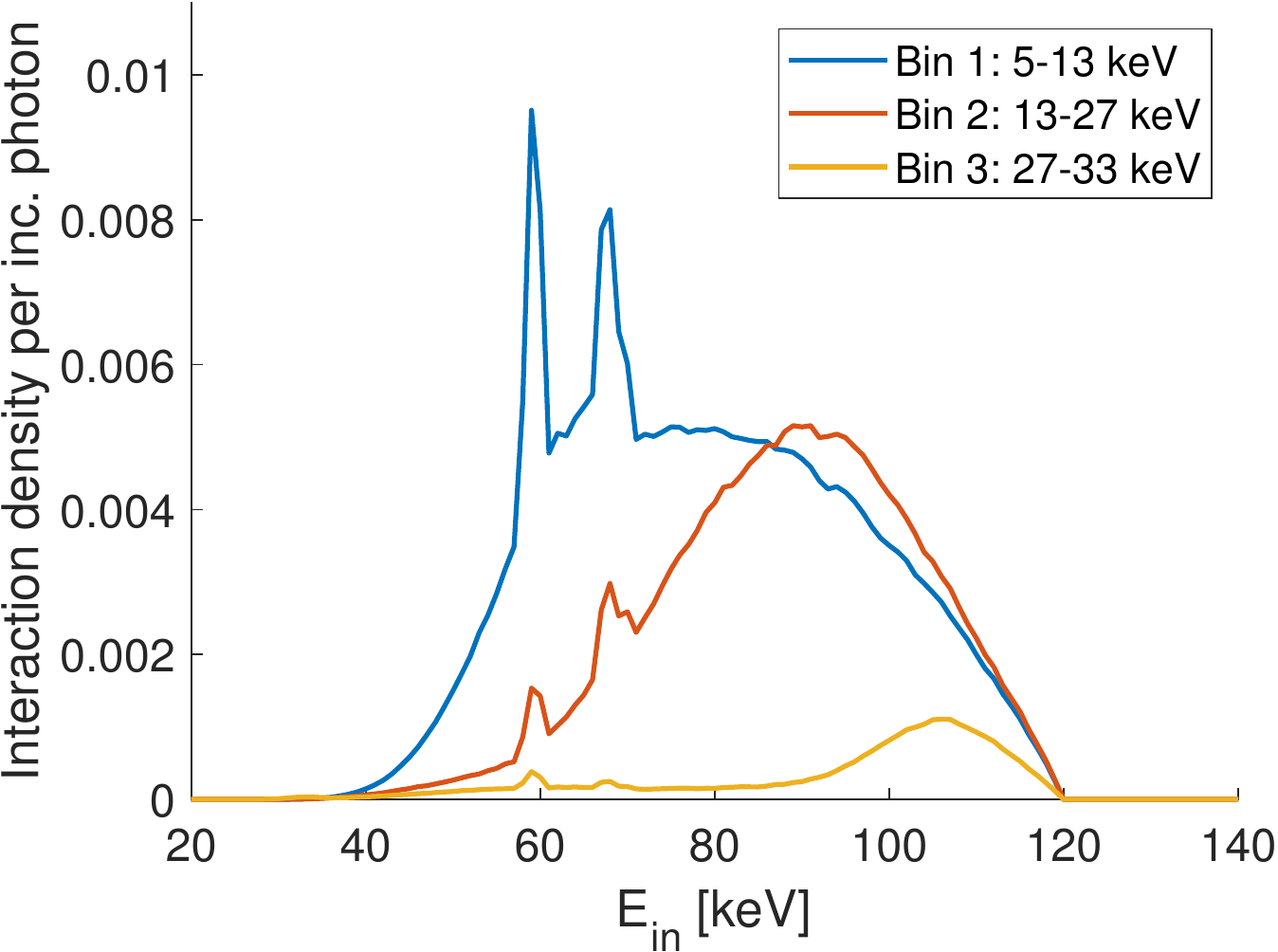}}
        \hfil
        \subfloat[]{\includegraphics[width=0.39\linewidth]{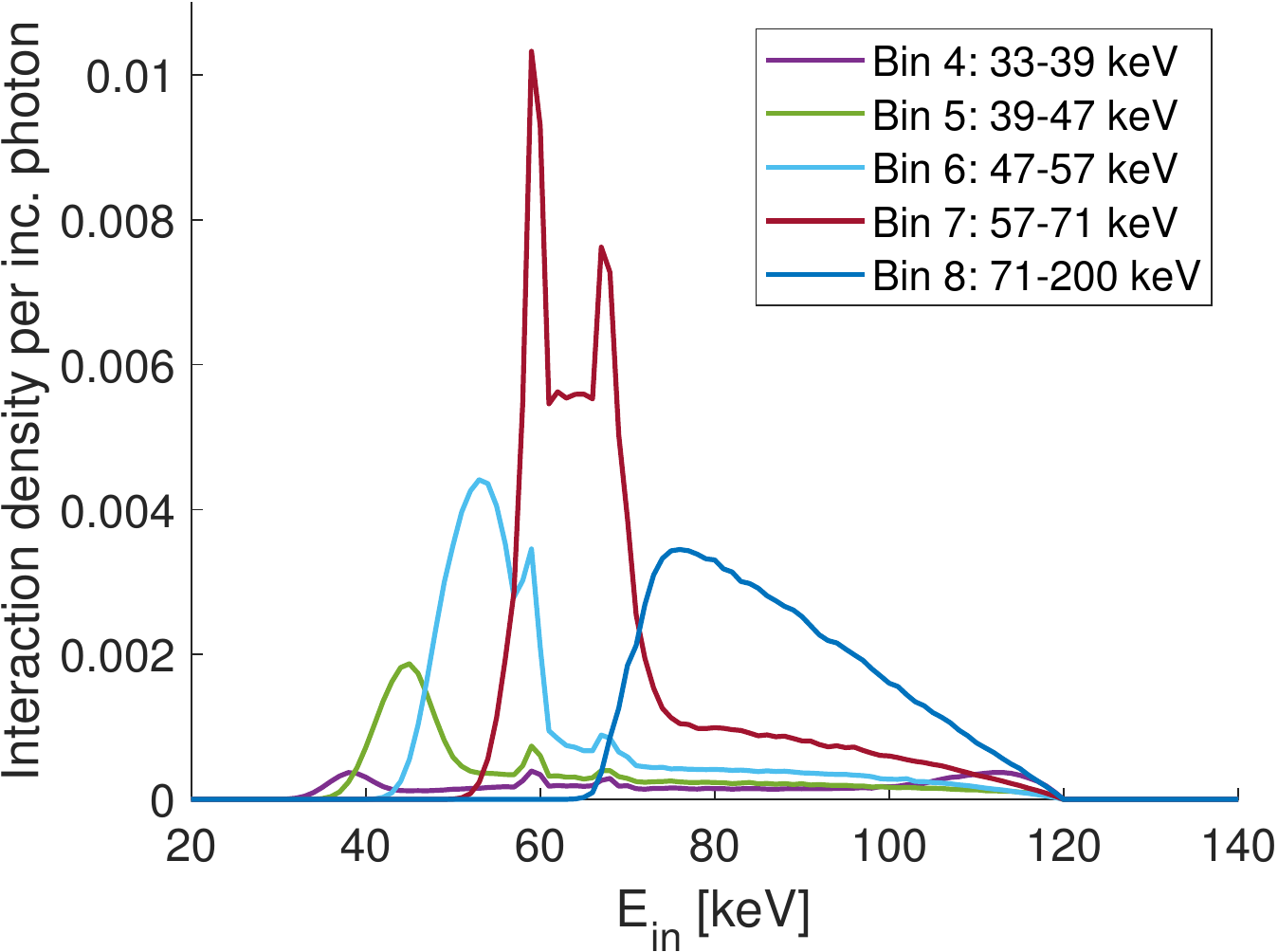}}
        \end{minipage}
        \caption{(a) Monoenergetic response of silicon showing distribution of photoelectric interactions, charge sharing and Compton scattering. (b) Incident and deposited spectra assuming a 120~kVp source spectrum and 30~cm water filtration. (c) Intrinsic bin response functions for lower energy bins. (d) Intrinsic bin response functions for higher energy bins. (e) Spectrum-weighted bin response functions for lower energy bins. (f) Spectrum weighted bin response functions for higher energy bins.}
        \label{fig:response_characteristics}
      \end{figure*}

      In Fig.\ref{fig:response_characteristics} (a) we show the large area response of the simulated \ds\ detector design to monoenergetic \ch{70~keV photons}. The characteristic features of the response are a photopeak, a charge sharing tail with a slight feature of \ch{K}-escape photons from the tungsten foils, and a Compton hump towards lower energies. 

      In Fig. \ref{fig:response_characteristics} (b) we show the incident, deposited and primary energy spectra for a large area assuming a 120~kVp  incident x-ray spectrum with 30~cm of water filtration. 
      
      As can be seen, the low energy part of the detected spectrum \ch{is comprised of} Compton interactions, \ch{which} are concentrated towards lower energies. \ch{A fraction of} photons \ch{in this energy range (the difference between the total deposited spectrum and the primary spectrum)} are photons that are counted multiple times. This implies that the lowest energy threshold can be used as an effective discriminator \ch{between the set containing both Compton events and multiple counting events, and the set of predominantly photopeak events.} \ch{Consequently, we can} study the impact of \ch{Compton scattering and multiple counting} in silicon by varying the lowest \ch{energy} threshold. Note that there are essentially no multiple counting events for thresholds above 35~keV.
  
      In Figs. \ref{fig:response_characteristics} (c) and (d) we show the intrinsic bin response functions of the detector, i.e. the probability that an incident photon will be counted in a given energy bin as a function of its energy, divided into low-energy bins and high-energy bins. In Figs. \ref{fig:response_characteristics} (e) and (f) the bin response functions have been weighted with the incident energy spectrum to form the probability densities for incident photons being counted in the respective energy bins. The high-energy bins show the expected behavior of being most sensitive to photons in the range of their thresholds, as well as having some sensitivity to higher energy photons due to charge sharing. The low-energy bins have a more complex behavior depending on the overlap of the Compton hump and energies in the incident spectrum. With the simulated threshold settings, the lowest energy bin is sensitive to photons in the entire range of the incident spectrum, whereas the second and third energy bins are mostly sensitive to higher energy photons. This shows that the energy deposited in a Compton interaction carries some information about the incident photon energy.

      The separability of the Compton and photoelectric parts of the deposited spectrum implies that the energy information in the photoelectric part of the spectrum is unaffected by the presence of Compton interactions. It is therefore possible to estimate the incident spectrum by deconvolving the measured spectrum with the spectral response \citep{bornefalk2010photon}. This is the reason why this type of non-overlapping spectral distortion tends to be less detrimental than distortion mechanisms that lead to overlap between misregistered and unaffected parts of the spectrum, such as charge sharing and K-fluorescence escape \cite{persson2020detective}.

    \subsubsection{The penalty of count multiplicity}\label{sec:count_multiplicity}

      The effect of internal detector scatter and cross-talk on zero-frequency DQE is well explained by considering count multiplicity. With internal detector scatter and cross-talk, some photons will be counted more than once. In this way, these photons contribute to extra signal, but also to extra variance. The effect on zero-frequency DQE is as follows. 

      Given an incident signal that is Poisson distributed, and a measured signal where incident events might be counted multiple times, zero-frequency DQE is reduced by a factor $\Ex[M]^2/\Ex[M^2]$, where $M$ is the random variable \ch{denoting} the number of times an incident photon gets counted. This factor is always less than or equal to one, with equality if $M$ is a constant. This can be shown using, for example, the laws of total expectation and variance as \ch{we now describe}.

      Let $X$ be a Poisson random variable with $\Var[X]=\Ex[X]$ and let $Y$ be the random variable $\sum_{i=1}^X M_i$, where $M_i$ is the random variable that is the number of times the $i$th photon \ch{is} counted, i.e., the count multiplicity. Since $M_1,\dots,M_X$ and $X$ are statistically independent from each other, the conditional mean and variance of $Y$ given $X$ are given by $\Ex[Y|X]=\Ex[M]\cdot X$ and $\Var[Y|X]=\Var[M]\cdot X$. Applying the law of total expectation then yields that 
      \begin{equation}\label{eq:multiplicity_mean}
        \Ex[Y]=\Ex[\Ex[Y|X]]=\Ex[M\cdot\Ex[X]]=\Ex[M]\cdot\Ex[X]      
      \end{equation}
      and applying the law of total variance yields:
      \begin{equation}\label{eq:multiplicity_variance}
        \begin{aligned}
        \Var[Y] &= \Ex[\Var[Y|X] + \Var[\Ex[Y|X]] \\
                &= \Ex[\Var[M]\cdot X] + \Var[\Ex[M]\cdot X] \\
                &= \Var[M]\cdot \Ex[X] + \Ex[M]^2\cdot \Var[X] \\
                &= \Var[M]\cdot\Ex[X] + \Ex[M]^2\cdot\Ex[X] \\
                &= (\Var[M] + \Ex[M]^2 )\cdot \Ex[X] \\
                &= \Ex[M^2]\cdot\Ex[X].
        \end{aligned}
      \end{equation}
      Now let $X_0$ and $X_1$ be Poisson variables corresponding to background and target respectively and let $Y_0$ and $Y_1$ denote $\sum_{i=1}^{X_k} M_i$ for both signals. By \eqref{eq:multiplicity_mean} and \eqref{eq:multiplicity_variance} it follows that the $\text{CNR}^2$ of $Y_0$ and $Y_1$ is given by
      \begin{equation}
        \text{CNR}_Y^2 = \frac{\left(\Ex[Y_1] - \Ex[Y_0] \right)^2}{\Var[Y_0]} = \frac{\Ex[M]^2\cdot\left(\Ex[X_1] - \Ex[X_0] \right)^2}{\Ex[M^2]\cdot\Ex[X_0]} =  \frac{\Ex[M]^2}{\Ex[M^2]}\cdot \text{CNR}_X^2
      \end{equation}
      and thus that
      \begin{equation}\label{eq:multiplicity_penalty}
        \text{DQE}_Y = \frac{\text{CNR}_Y^2}{\text{CNR}_X^2} = \frac{\Ex[M]^2}{\Ex[M^2]} \leq 1.
      \end{equation}    
      This quantity should be interpreted as a factor corresponding to the reduction in DQE due to multiple counting. To obtain the total DQE of a detector this should be multiplied with other factors that reduce DQE, such as quantum detection efficieny.

  \subsubsection{Geometric Poisson model for scatter}\label{sec:geometric_poisson}

    A simple model for the effect of Compton scatter on zero-frequency DQE can be obtained by assuming that the number of times a photon is counted is given by a geometric random variable. This will make the total number of counts in a measurement a geometric Poisson random variable. This corresponds to assuming that each photon has a constant probability $\rho$ of scattering and being counted again, independent of its energy and how many times it has been counted already. The count multiplicity $M$ then has a probability \ch{density} function $\Prob(M=m) = \rho^{m-1}(1 - \rho)$, with mean and variance given by
    \begin{equation}\label{eq:geometric_mean}
      \Ex[M] = \frac{1}{1-\rho},\quad \Var[M] = \frac{\rho}{(1-\rho)^2}.    
    \end{equation}
    It follows that 
    \begin{equation}
      \Ex[M^2] = \Var[M] + \Ex[M]^2 = \frac{1 + \rho}{(1-\rho)^2},
    \end{equation}
    and that the zero-frequency DQE penalty, given by \eqref{eq:multiplicity_penalty}, is
    \begin{equation}
      \frac{\Ex[M]^2}{\Ex[M^2]} = \frac{1}{1+\rho}.
    \end{equation}
    To obtain an estimate of $\rho$ we relate the mean of $M$ to the scatter-to-primary ratio (SPR) \ch{for scatter that occurs within the detector}:
    \begin{equation}
      \frac{1}{1-\rho} = 1 + \text{SPR},
    \end{equation}
    which corresponds to assuming that $\rho$ is equal to $\text{SPR}/(1+\text{SPR})$, or the scatter-to-total ratio. Expressed in terms of SPR, the zero-frequency DQE penalty can the be shown to equal
    \begin{equation}\label{eq:geometric_dqe_penalty}
      \frac{1}{1+\rho} = \frac{1+\text{SPR}}{1+2\,\text{SPR}}.
    \end{equation}
    We will investigate the validity of this model in later sections. 

  \subsection{Detector simulation}\label{sec:Detector_Simulation}

    The response of an \ds\ detector \ch{were} simulated \ch{using} a Monte Carlo model described in \cite{persson2020detective}. In brief, photon transport in the detector material was simulated with the Geant4 Application for Emission Tomography (GATE) \cite{jan2004gate}, and a three-dimensional Gaussian charge cloud model ($\sigma=2.0\cdot E^{0.53}$~$\mu$m) was used to simulate the effect of charge sharing. Incident monoenergetic beams with energies of 20--150 keV in steps of 1~keV, with 25000 photons for each energy, were simulated. The beam covered either one pixel unit cell (for simulating the ACF) or $1/3 \times 1/3$ unit \ch{cells} (for simulating the PSF). The effect\ch{s} of pileup, scatter from the imaged object, \ch{and} an anti-scatter grid were not included in the model. For each simulated incident energy, the total number of counts in each detector pixel was recorded at each deposited energy level in 1~keV steps. Furthermore, for each lowest threshold position (in 1~keV steps), the number of unique photons leading to at least one recorded count above this energy was recorded. \ch{The} difference between the total number of counts, \ch{and the number of} unique counts corresponds to the number \ch{of} additional counts from multiply-counted photons. This will be referred to as scatter, and we take the ratio of scatter and unique counts as the SPR. 

    The unbinned spatio-energetic pixel PSF, $h^{\text{ub}}_{\boldsymbol{n}}(E,E')$, which gives the expected number of counts registered with energy $E'$ in the pixel with index $\boldsymbol{n}=(n_x,n_y)$ for an incident beam with energy $E$ covering the central pixel with index $(0,0)$, is obtained \ch{by} computing the sample average of the Monte Carlo data and adding the sub-pixel PSFs. \ch{By computing the variance of the Monte Carlo data when an incident beam with energy $E$ covers the central pixel, we obtain the} unbinned pixel ACF, $K^{\text{ub}}_{\boldsymbol{n}}(E,E',E'')$. \ch{This is} the covariance of the number of counts registered with energy $E'$ in the central pixel \ch{having} index $(0,0)$, \ch{in the case of an} energy $E''$ \ch{being recorded} in the pixel with index $\boldsymbol{n}$. 

    The effect of electronic readout noise was added to the point-spread function and autocovariance function by convolving \ch{each} with a Gaussian kernel with an RMS of 1.6~keV \ch{along} the \ch{dimension} corresponding to detected energy. For the case $\boldsymbol{n} = 0$, the autocovariance is convolved only along its diagonal, since an interaction can only be counted once in a single pixel and \ch{cannot} therefore give rise to any correlation in the case of Poisson statistics. For a given set of energy thresholds, $T_1,T_M,\dots,T_M$ the binned spatio-energetic pixel PSF and ACF, \ch{respectively} denoted $h_{\boldsymbol{n},j}(E)$ and $K_{\boldsymbol{n},j,j'}(E)$, where the indices $j$ and $j'$ refer to energy bins, are then obtained by integrating over the energy ranges defined by the thresholds. Details of this computation \ch{appear} in Appendix \ref{sec:appendix_a}. 

    The binned large-area PSF and ACF are then obtained by summing over pixel indices:
    \begin{equation}
      h_{j}(E) = \sum_{n_x=-\infty}^{\infty}\, \, \sum_{n_y=-\infty}^{\infty}h_{\boldsymbol{n},j}(E), 
    \end{equation}
    and
    \begin{equation}
      K_{j,j'}(E) = \sum_{n_x=-\infty}^{\infty}\,\,\sum_{n_y=-\infty}^{\infty}K_{\boldsymbol{n},j,j'}(E).
    \end{equation}
    We similarly obtain the binned large-row PSF and ACF, which we \ch{respectively} denote as $h_{n,j}(E)$ and $K_{n,j,j'}(E)$, by summing only over $n_y$.

  \subsection{Geometric-Poisson model validation}\label{sec:geometric_poisson_validation}
    
    To validate the Geometric-Poisson model developed in Section \ref{sec:geometric_poisson}, we use the SPR obtained from the Monte Carlo simulation to compute the DQE penalty predicted by \eqref{eq:geometric_dqe_penalty} as a function of the lowest threshold. We also use the binned large-area PSF and ACF to compute the DQE predicted by the Monte Carlo simulation. For both models we assume a purely photon-counting detector with only one threshold. 

  \subsection{Projection domain evalutation}\label{sec:projection_evaluation}

    To evaluate the effect of Compton scatter in the projection domain, we use the binned large area PSF and ACF to compute the DQE for two particular imaging tasks given a 120 kVp background spectrum with 30~cm of water filtration. The tasks correspond to those material combinations that are the \emph{easiest} and the \emph{hardest} to distinguish from the statistical noise in the background. The easiest task corresponds to a pure density imaging task in the sense that there is minimal spectral information available, as the underlying pathlength contrast induces a relative change in the bin counts that is approximately equal for all energy bins. The hardest task, on the other hand, corresponds to a pure spectral task, for which the underlying pathlength contrast induces a change only in the distribution of counts across energy bins and no change in the total number of counts. This spectral task is thus impossible without an energy resolving detector. Note that these tasks have a slight dependence on the detector design and the lowest threshold position.

    Assuming that the incident spectrum is given by $q(E;\boldsymbol{A})$ where $\boldsymbol{A}=(A_1, A_2)$ is a pathlength vector corresponding to two basis materials, e.g., water and bone, the large-area bin response and covariance are given by
    \begin{equation}
      \lambda_j(\boldsymbol{A}) = \int_\mathbb{R} h_{j}(E)\,q(E;\boldsymbol{A})\,dE
    \end{equation}
    and 
    \begin{equation}
      \Sigma_{j,j'}(\boldsymbol{A}) = \int_\mathbb{R} K_{j,j'}(E)\,q(E;\boldsymbol{A})\,dE.
    \end{equation}
    Furthermore the partial derivative of the response with respect to a basis material \ch{are} given by 
    \begin{equation}
      \frac{\partial\lambda_j(\boldsymbol{A})}{\partial A_l} = \int_\mathbb{R} h_{j}(E)\,\,\frac{\partial q(E;\boldsymbol{A})}{\partial A_l}\, dE.
    \end{equation}

    The Cramer-Rao lower bound (CRLB) matrix $\boldsymbol{C}$ \ch{(the lower bound on the covariance of any unbiased estimator)} of the basis material pathlength vector $\boldsymbol{A}$, is given by:
    \begin{equation}
      \boldsymbol{C} = \left(\left(\frac{\partial\boldsymbol{\lambda}}{\partial \boldsymbol{A}}\right)\boldsymbol{\Sigma}^{-1}\left(\frac{\partial\boldsymbol{\lambda}}{\partial \boldsymbol{A}}\right)^\top\right)^{-1},
    \end{equation}
    where $\big[\frac{\partial\boldsymbol{\lambda}}{\partial \boldsymbol{A}}\big]_{j,l} = \frac{\partial\lambda_j(\boldsymbol{A})}{\partial A_l}$ and $\big[\boldsymbol{\Sigma}\big]_{j,j'} = \Sigma_{j,j'}(\boldsymbol{A})$ denote the mean vector and the covariance matrix of registered counts, \ch{respectively}. For an ideal photon-counting detector, the CRLB matrix $\boldsymbol{C}^{\text{ideal}}$ is obtained in the same way by assuming that the PSF and ACF are given by Dirac delta functions. 

    Let $\boldsymbol{v}_{\text{min}}$ and $\boldsymbol{v}_{\text{max}}$ denote the eigenvectors corresponding to the smallest and \ch{largest} eigenvalues $\sigma^2_{\text{min}}$ and $\sigma^2_{\text{max}}$ of $\boldsymbol{C}$. These vectors correspond exactly to the easiest and hardest imaging task\ch{s} described above. The large-area projection-domain DQE for the density and spectral imaging tasks are then defined as:
    \begin{equation}
      \text{DQE}^{\,\text{density}}_{\,\text{projection}} = \frac{\boldsymbol{v}_{\text{min}}^\top \boldsymbol{C}^{\text{\,ideal}}\boldsymbol{v}_{\text{min}}}{\boldsymbol{v}_{\text{min}}^\top \boldsymbol{C} \boldsymbol{v}_{\text{min}}}, \quad \text{DQE}^{\,\text{spectral}}_{\,\text{projection}} = \frac{\boldsymbol{v}_{\text{max}}^\top \, \boldsymbol{C}^{\text{\,ideal}}\boldsymbol{v}_{\text{max}}}{\boldsymbol{v}_{\text{max}}^\top \, \boldsymbol{C} \boldsymbol{v}_{\text{max}}},
    \end{equation}
    where we recognize that the quadratic form $\boldsymbol{v}^\top\boldsymbol{C}\boldsymbol{v}$ is the optimal $\text{CNR}^2$ for the task corresponding to the contrast defined by $\boldsymbol{v}$ \cite{tapiovaara1985snr,schmidt2009optimal}. Similarly, the DQE for a monoenergetic image at a given energy is defined using the weights used to form that image. 

    We investigate the effect of Compton scatter on these metrics by varying the lowest energy threshold from zero keV to the maximum energy in the incident spectrum. We consider two sets of energy thresholds. First, in order to eliminate the effect of optimal bin selection, we assume that the thresholds are given by $T_1,T_1+1,\dots,E_\text{max}$, where $T_1$ is the lowest threshold and $E_\text{max}$ is the maximum energy in the detected spectrum. Second, for a more realistic scenario we assume that the detector has 8 energy thresholds given by $T_j = \max(T_j^{\,0},T_1)$, where $T_j^{\,0}$ denotes a nominal threshold and $T_1$ is the lowest threshold. The nominal thresholds were found by optimizing the sum of $\text{DQE}^{\text{density}}_{\text{projection}}$ and $\text{DQE}^{\,\text{spectral}}_{\,\text{projection}}$ for backgrounds from 0--50~cm of water in \ch{integral} cm steps with the lowest threshold fixed at 5~keV. The resulting thresholds were given by 5, 13, 29, 34, 39, 48, 56 and 73~keV. Note that \ch{these} thresholds will be sub-optimal for any other lower threshold than 5~keV, and also that the number of energy bins decreases whenever the lowest threshold crosses one of the nominal thresholds. This setup was nonetheless chosen for having more easily interpretable results, with less sensitivity to threshold optimization. 

  \subsection{Image domain evaluation}\label{sec:image_evaluation}

    We compute analogous DQE metrics in the image domain by simulating CT imaging of a cylindrical 30~cm water phantom, but instead of the CRLB, we use the ensemble covariance of basis material image noise:
    \begin{equation}
        \text{DQE}^{\,\text{density}}_{\text{\,image}} = \frac{\boldsymbol{v}_{\text{min}}^\top \widehat{\boldsymbol{C}}^{\text{\,ideal}}\boldsymbol{v}_{\text{min}}}{\boldsymbol{v}_{\text{min}}^\top \widehat{\boldsymbol{C}} \boldsymbol{v}_{\text{min}}}, \quad \text{DQE}^{\text{\,spectral}}_{\text{\,image}} = \frac{\boldsymbol{v}_{\text{max}}^\top \widehat{\boldsymbol{C}}^{\text{\,ideal}}\boldsymbol{v}_{\text{max}}}{\boldsymbol{v}_{\text{max}}^\top \widehat{\boldsymbol{C}} \boldsymbol{v}_{\text{max}}},
        \end{equation}
    where $\widehat{\boldsymbol{C}}$ and $\widehat{\boldsymbol{C}}^{\text{\,ideal}}$ are the sample ensemble covariance matrices of reconstructed basis image values for the simulated silicon detector and an ideal photon-counting detector, respectively, and $\boldsymbol{v}_{\text{min}}$ and $\boldsymbol{v}_{\text{max}}$ are the eigenvectors corresponding to the smallest and greatest eigenvalue\ch{s} of $\widehat{\boldsymbol{C}}$. The sample ensemble covariance matrices are computed as follows, similar to the methodology in \cite{tong2010noise}. For a set of $K = 60$ non-overlapping $15\times15$ $\text{mm}^2$ regions of interest (ROIs) spread out over the whole phantom in the reconstructed images and $N = 50$ ensemble realizations. \ch{L}et $\boldsymbol{a}_{k,n}$ denote the vector of average values in each basis image of the $k$th ROI in the $n$th ensemble realization, and let $\bar{\boldsymbol{a}}_k$ denote the ensemble average of $\boldsymbol{a}_{k,n}$. The sample ensemble covariance $\widehat{\boldsymbol{C}}$ is then:
    \begin{equation}
      \widehat{\boldsymbol{C}} = \frac{1}{K}\sum_{k=1}^{K} \widehat{\boldsymbol{C}}_k, \quad \widehat{\boldsymbol{C}}_k = \frac{1}{N-1}\sum_{n=1}^{N} (\boldsymbol{a}_{k,n} - \bar{\boldsymbol{a}}_k)(\boldsymbol{a}_{k,n} - \bar{\boldsymbol{a}}_k)^\top.
    \end{equation}

    We simulate imaging of a 30~cm \ch{cylindrical water phantom} in a fan beam geometry with a source-to-isocenter distance of 50~cm and a source-to-detector distance of 100~cm. We use a 120~kVp tube spectrum and assume a total current-time product of 200~mAs and 2000 views over a full rotation. We assume a large row setting, corresponding to a 10~mm slice \ch{at} isocenter. To simplify the addition of correlated noise, we furthermore assume an ideal bowtie filter in the sense that it gives a flat field of illumination with the object present in the beam. Let $\boldsymbol{A}_n$ denote the simulated ground truth pathlength vector in front of pixel $n$ and let $q_n(E) = q(E;\boldsymbol{A}_n)$ denote the simulated incident spectrum on pixel $n$. For the ideal detector we assume that the number of photons detected with energy $E$ is a Poisson random variable with mean $q_n(E)$. For the silicon detector we compute the large-row response and autocovariance function for 8 energy bins given by $T_j = \max(T_j^{\,0},T_1)$, where $T_j^{\,0}$ denotes the nominal threshold and $T_1$ is varied from $0$ to $30$~keV. The same nominal thresholds were used as for the projection domain evaluation. 

    The average signal level in each bin of pixel $n$ is computed by convolving $q_n$ with the large row response function:
    \begin{equation}
        \lambda_{n,j} = \sum_{k=-\infty}^\infty \int_\mathbb{R} h_{k,j}(E)\,\,q_{n-k}(E)\,dE,
    \end{equation}
    and similarly the spatio-energetic covariance is computed by integrating $q_n$ with the large row autocovariance function
    \begin{equation}
        \Sigma_{n,n',j,j'} = \int_\mathbb{R} K_{n',j,j'}(E)\,\,q_n(E)\,dE.
    \end{equation}
    The detected signal for the silicon detector is then assumed to be a normal random variable with mean $\boldsymbol{\lambda}$ and covariance $\boldsymbol{\Sigma}$, where $\boldsymbol{\lambda}$ is the vector representation of $\lambda_{n,j}$ and $\boldsymbol{\Sigma}$ the corresponding covariance matrix. We simulate 50 independent \ch{count data} realizations for each setting of the lowest threshold in order to compute ensemble statistics.

    Let $Y_{n,j}$ denote the simulated number of counts with added statistical noise in pixel $n$ and bin $j$. The projection-based maximum likelihood estimator of the pathlength vector $\boldsymbol{A}_n$, assuming Poisson statistics, is given by 
    \begin{equation}\label{eq:ML_MD}
      \boldsymbol{A}_n^* = \argmin_{\boldsymbol{A}} \sum_{j=1}^{M}\lambda_{n,j}(\boldsymbol{A}) - Y_{n,j}\log\lambda_{n,j}(\boldsymbol{A}).
    \end{equation}
    In order to compute material-decomposed sinograms we use a linearized version of the maximum-likelihood estimator obtained by linearization around the simulated ground truth $\boldsymbol{A}_n$:
    \begin{equation}\label{eq:lin_mle}
      \boldsymbol{A}_n^* = \boldsymbol{A}_n + \left(\boldsymbol{J}_n^\top\boldsymbol{\Lambda}_n^{-1}\boldsymbol{J}_n\right)^{-1}\boldsymbol{J}_n^\top\boldsymbol{\Lambda}_n^{-1}\left(\boldsymbol{Y}_n - \boldsymbol{\lambda}_n\right),
    \end{equation}
    where $\big[\boldsymbol{J}_n\big]_{jk} = \frac{\partial\lambda_{n,j}}{\partial A_k}$ is the Jacobian of the forward model evaluated at $\boldsymbol{A}_n$ and $\boldsymbol{\Lambda}_n = \text{diag}(\boldsymbol{\lambda}_n)$ is the diagonal matrix of the forward model evaluated at $\boldsymbol{A}_n$. This estimator is unbiased and efficient in the case of Poisson statistics, which can be shown by computing the expectation and covariance of the expression in \eqref{eq:lin_mle}. Note that this estimator is not implementable in practice because it requires knowledge of the ground truth pathlengths $\boldsymbol{A}_n$. It is, however, much faster than solving \eqref{eq:ML_MD} and has the same variance as an optimal unbiased estimator would. 

    The estimated pathlengths for each view were reconstructed into basis images using Matlab's \texttt{ifanbeam} function with an image grid size of 1024 pixels, using a Hann kernel with unit frequency scaling and linear interpolation.

\section{Results}\label{sec:Results}

\begin{figure*}
      \centering
      \begin{minipage}[b]{\linewidth}
      \subfloat[]{\includegraphics[width=0.4\linewidth]{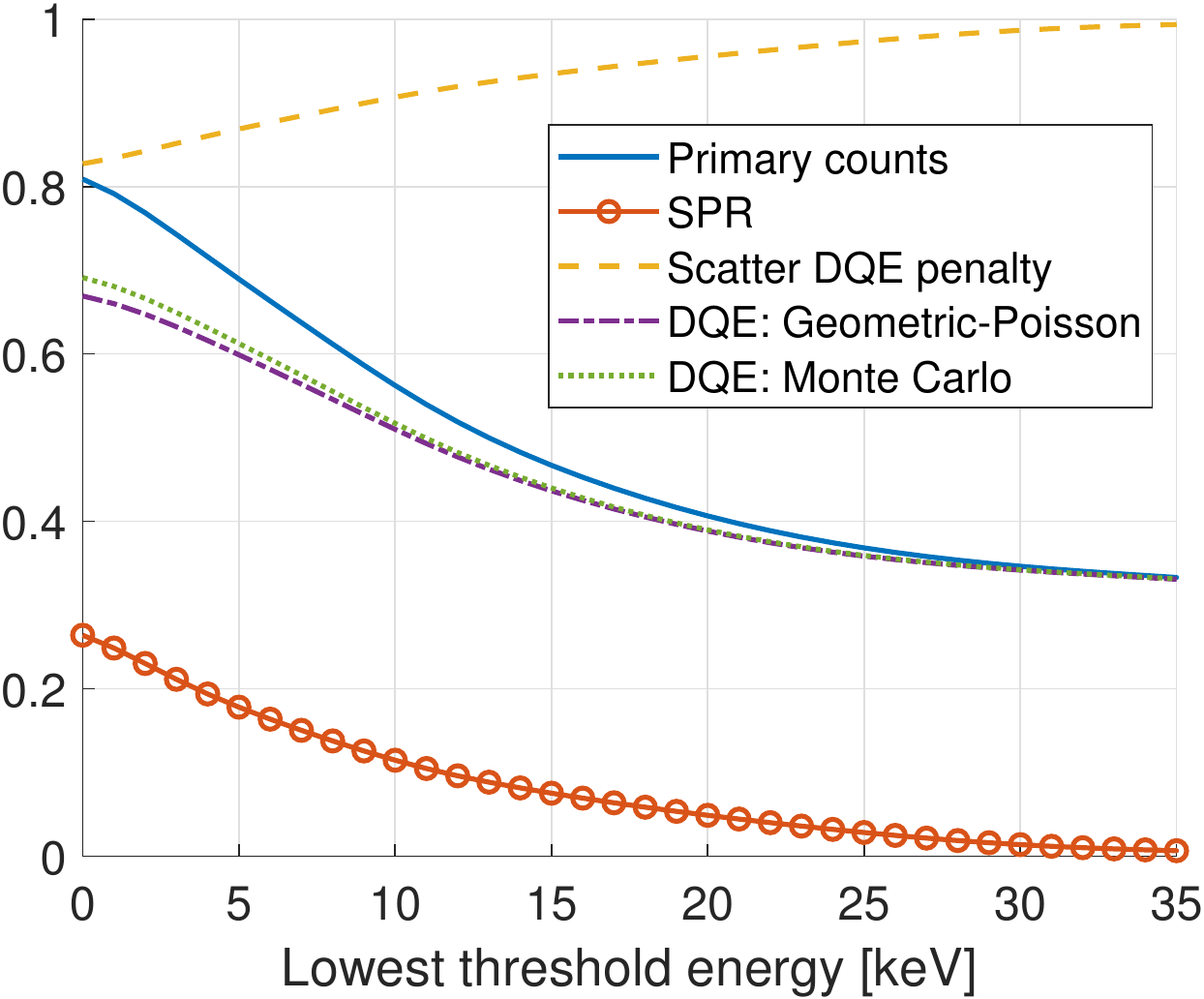}}        
      \end{minipage}
      \caption{Comparison of \ch{the} DQE penalty of multiple count \ch{events} predicted by the Geometric-Poisson model and Monte Carlo simulation. The different curves shown are: the number of primary counts above the lowest threshold (solid, blue), the scatter-to-primary ratio (solid, red, rings), the DQE penalty predicted by the Geometric-Poisson model (dashed, yellow), the DQE predicted by the Geometric-Poisson model, i.e., the product of the number of primary counts and the predicted DQE penalty (dash-dotted, purple) and the DQE obtained from Monte Carlo simulations (dotted, green). }
      \label{fig:dqe_penalty_breakdown}
  \end{figure*}

\begin{figure*}[htpb]
        \centering
        \begin{minipage}[b]{\linewidth}
        \subfloat[]{\includegraphics[width=0.4\linewidth]{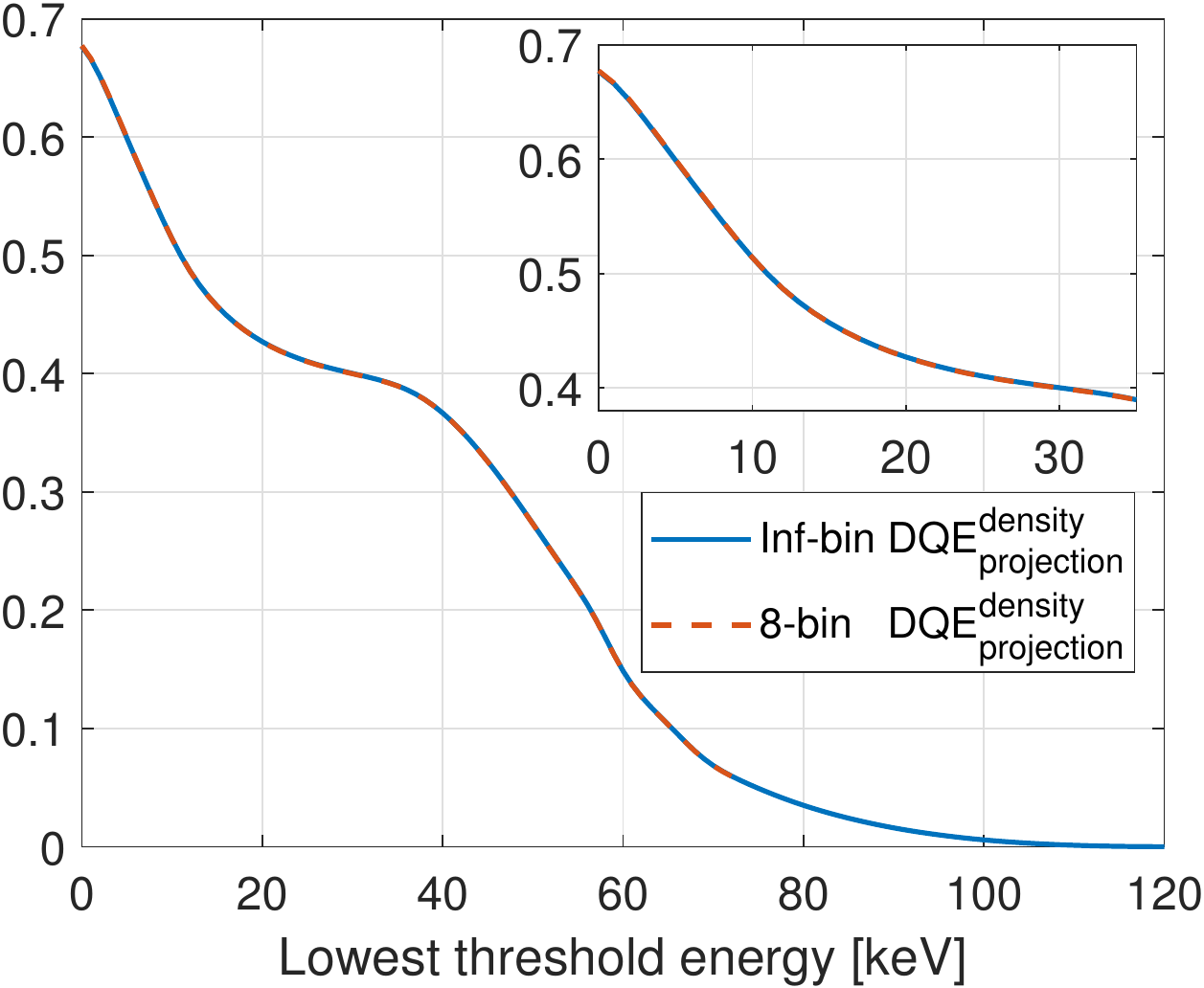}}
        \hfil
        \subfloat[]{\includegraphics[width=0.4\linewidth]{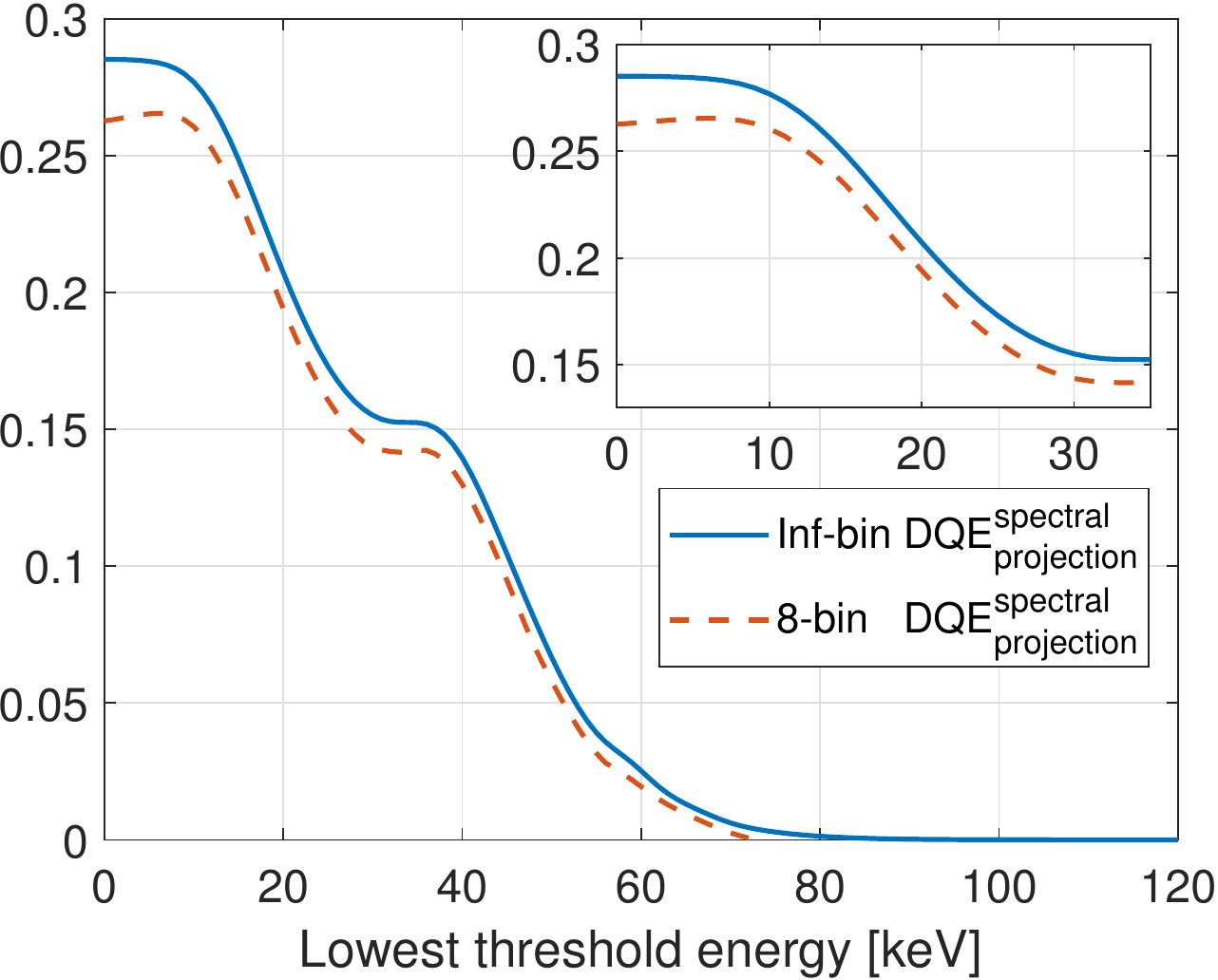}}    
        \vfil
        \subfloat[]{\includegraphics[width=0.4\linewidth]{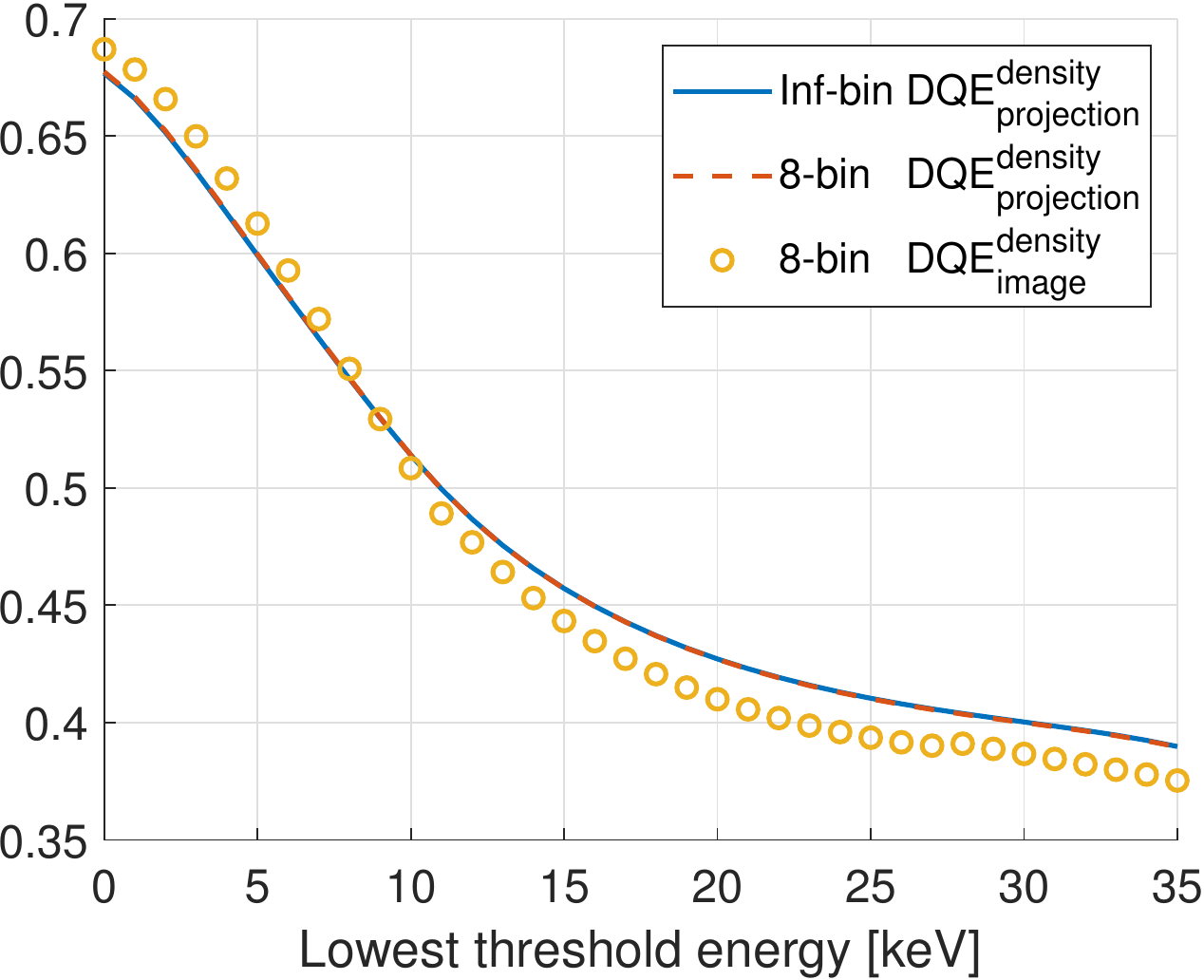}}
        \hfil
        \subfloat[]{\includegraphics[width=0.4\linewidth]{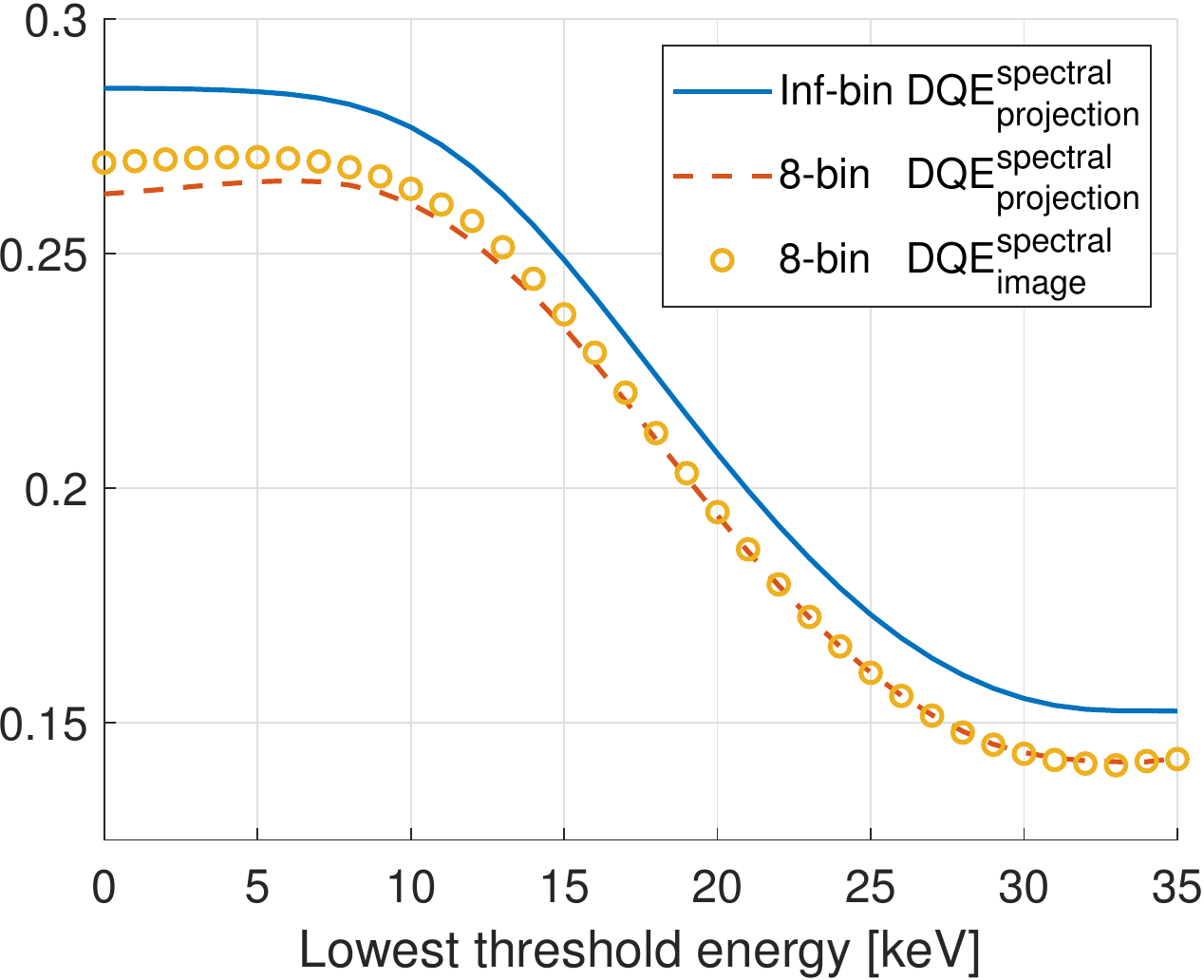}}  
        \end{minipage}
        \caption{(a) Density (easiest task) DQE in the projection domain assuming a 120~kVp source spectrum and 30~cm water filtration, assuming 120 energy bins with \ch{integral} spacing (solid, blue) and 8 energy bins (dashed, red). (b) Spectral (hardest task) DQE in the projection domain assuming a 120~kVp source spectrum and 30~cm water filtration, assuming 120 energy bins with \ch{integral} spacing (solid, blue) and 8 energy bins (dashed, red). (c) Density DQE in the image domain for a 30~cm water phantom imaged with a 120~kVp source spectrum, assuming 8 energy bins (yellow, circles), in comparison with density DQE in the projection domain from (a). (d) Spectral DQE in the image domain for a 30~cm water phantom imaged with a 120~kVp source spectrum, assuming 8 energy bins (yellow, circles), in comparison with spectral DQE in the projection domain from (b).}
        \label{fig:dqe}
  \end{figure*}

\begin{figure*}[htpb]
        \centering
        \begin{minipage}[b]{\linewidth}
        \subfloat[]{\includegraphics[width=0.35\linewidth]{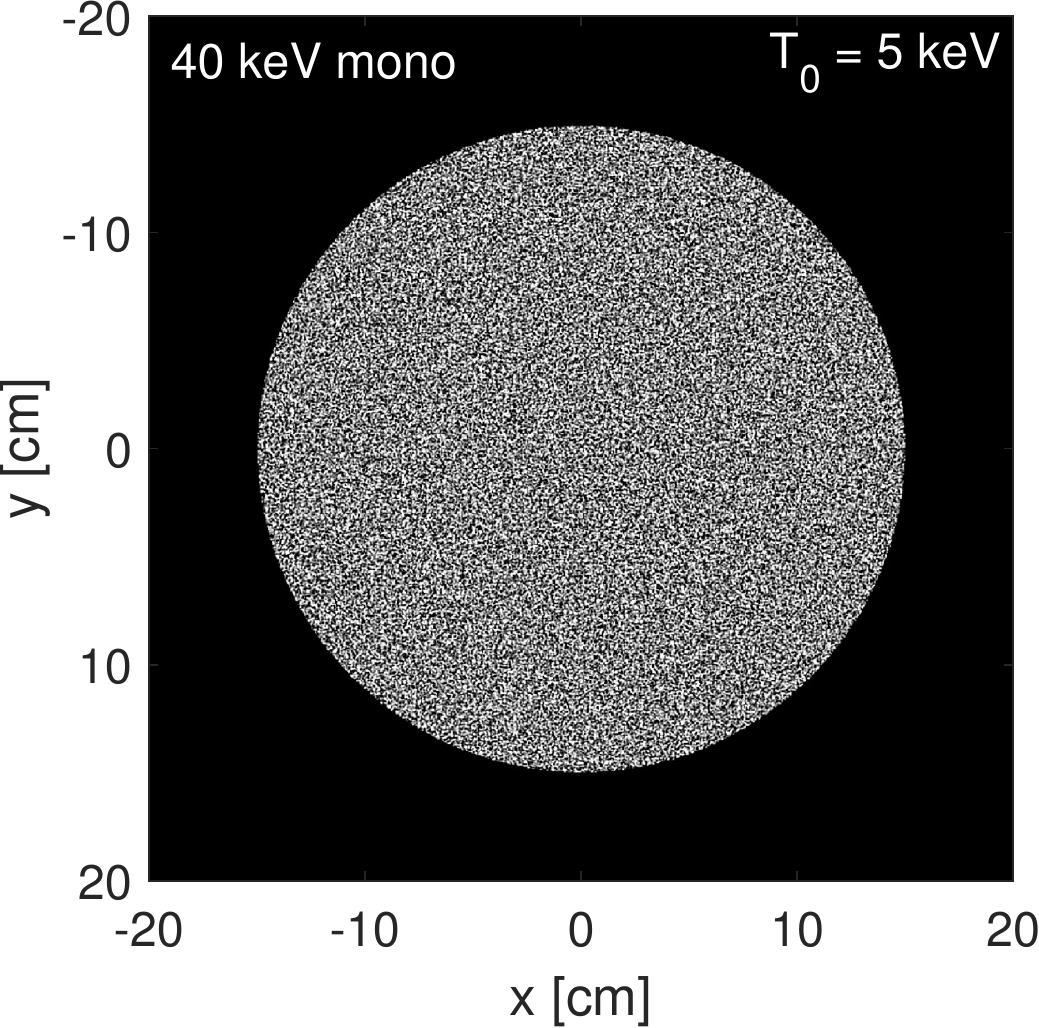}}
        \hfil
        \subfloat[]{\includegraphics[width=0.35\linewidth]{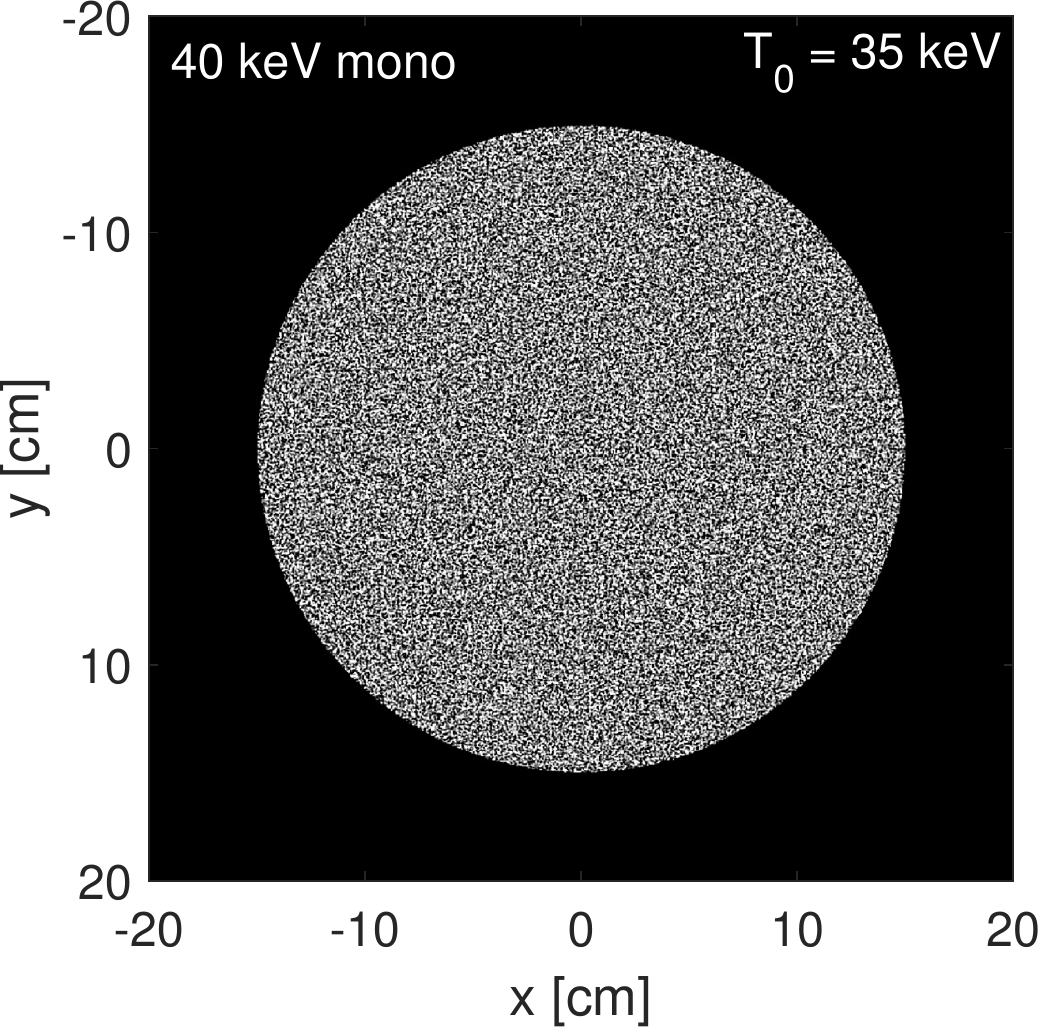}}   
        \vfil
        \subfloat[]{\includegraphics[width=0.35\linewidth]{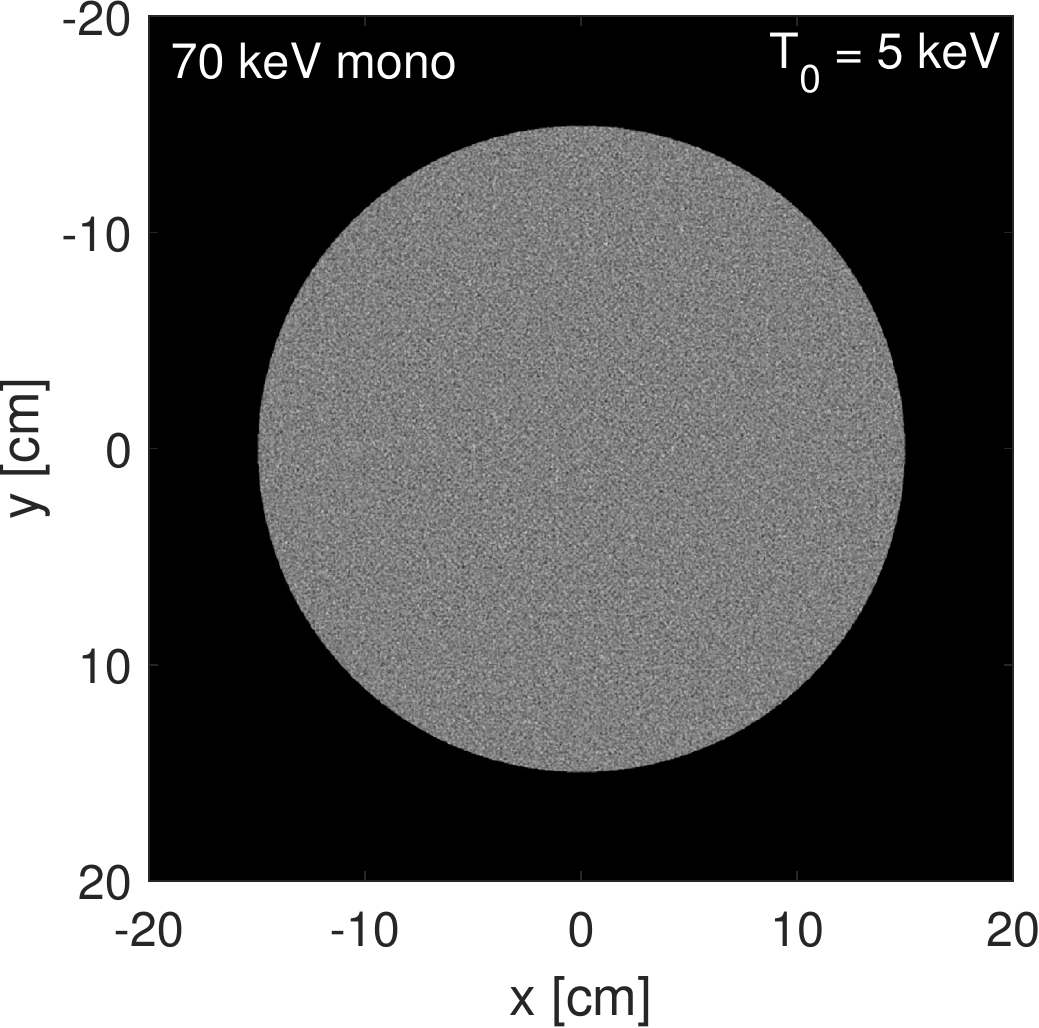}}
        \hfil
        \subfloat[]{\includegraphics[width=0.35\linewidth]{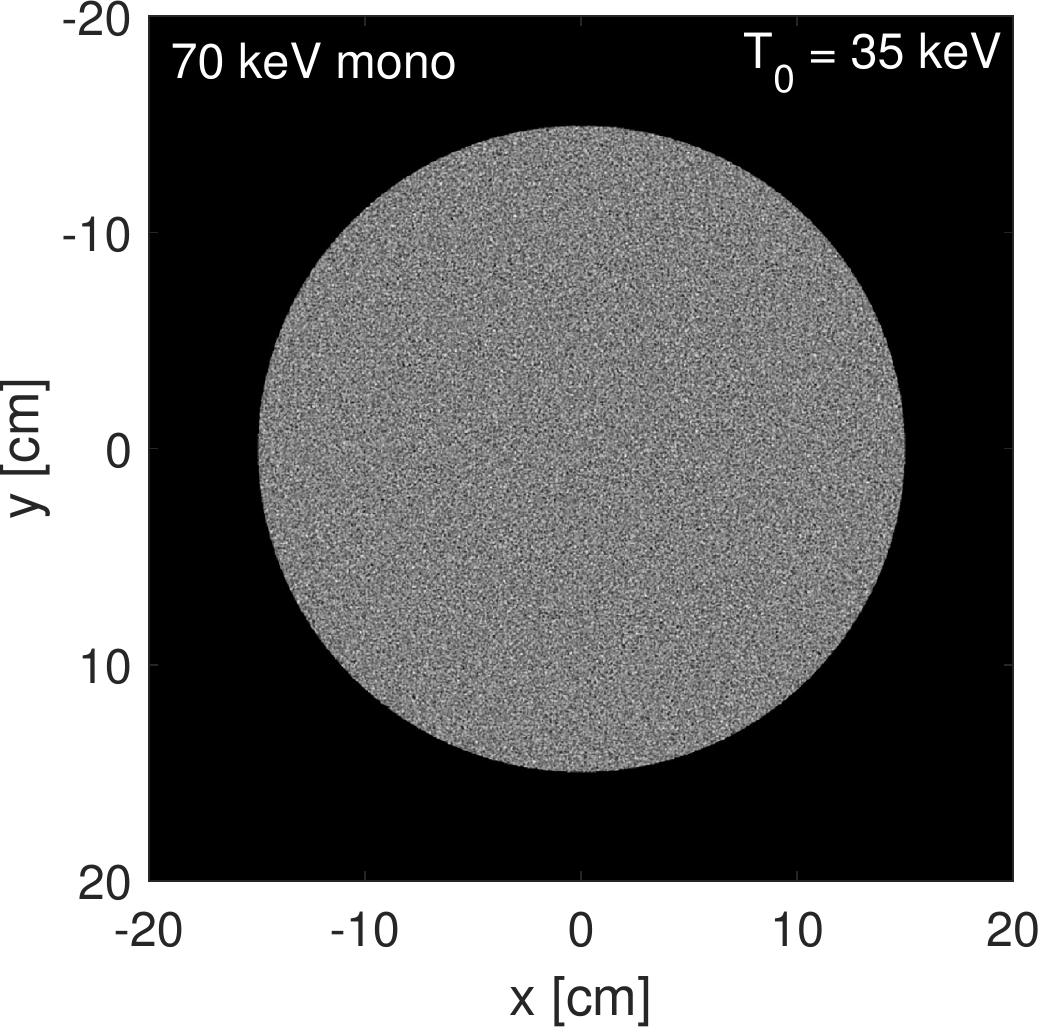}}   
        \vfil
        \subfloat[]{\includegraphics[width=0.35\linewidth]{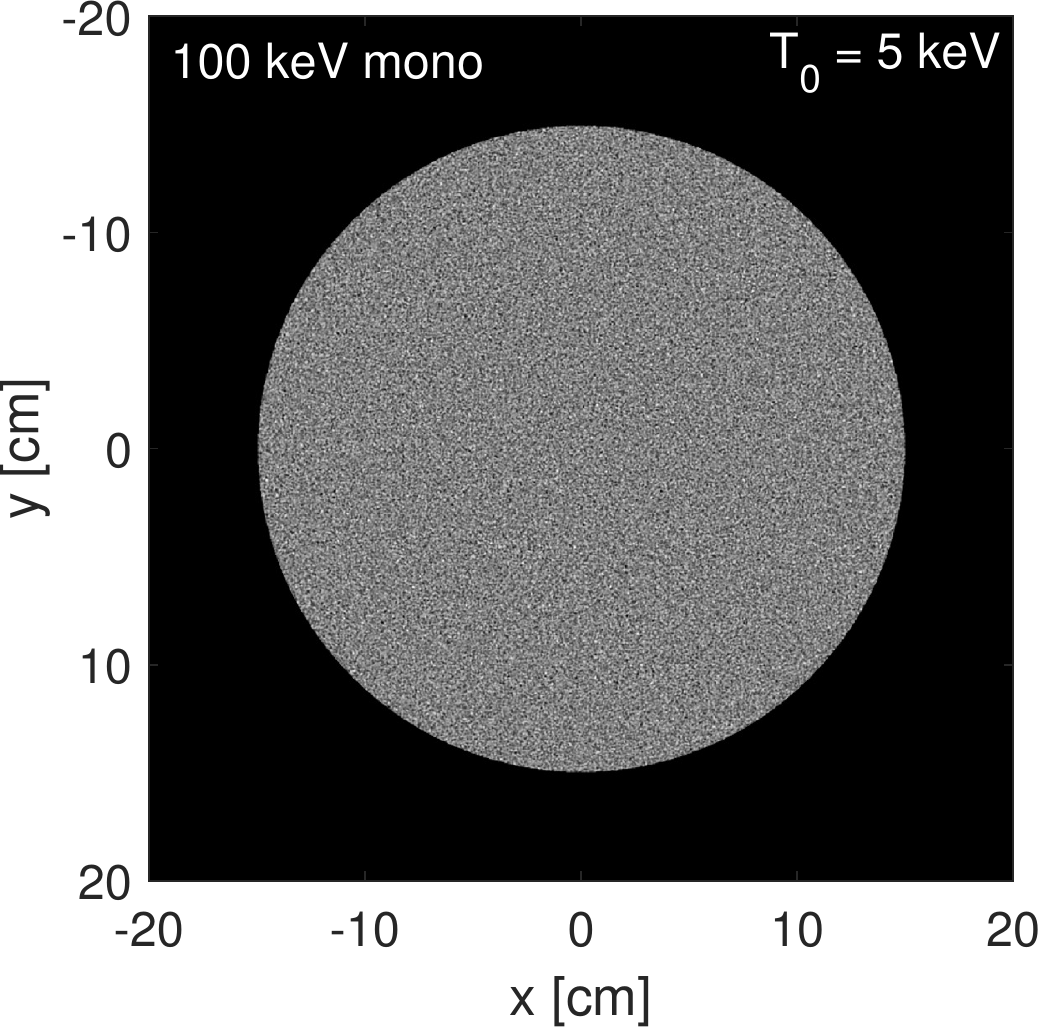}}
        \hfil
        \subfloat[]{\includegraphics[width=0.35\linewidth]{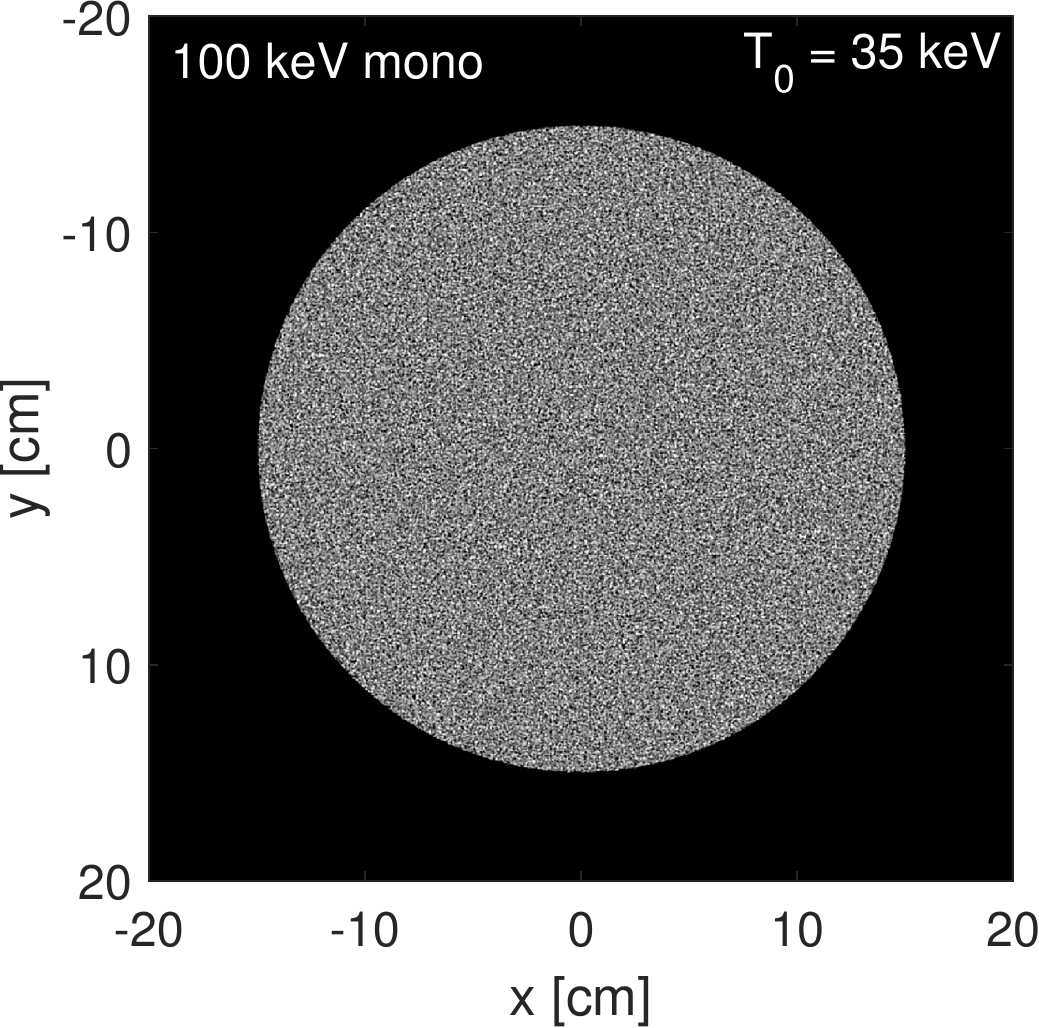}}             
        \end{minipage}
        \caption{Synthetic monoenergetic reconstructions at 40~keV (top row), 70~keV (middle row) and 100~keV (bottom row) from the simulated 8-bin \ds\ detector with lowest thresholds of 5~keV (left column) and 35~keV (right column).}
        \label{fig:mono_images}
  \end{figure*}

\begin{figure*}[htpb]
        \centering
        \begin{minipage}[b]{\linewidth}      
        \subfloat[]{\includegraphics[width=0.4\linewidth]{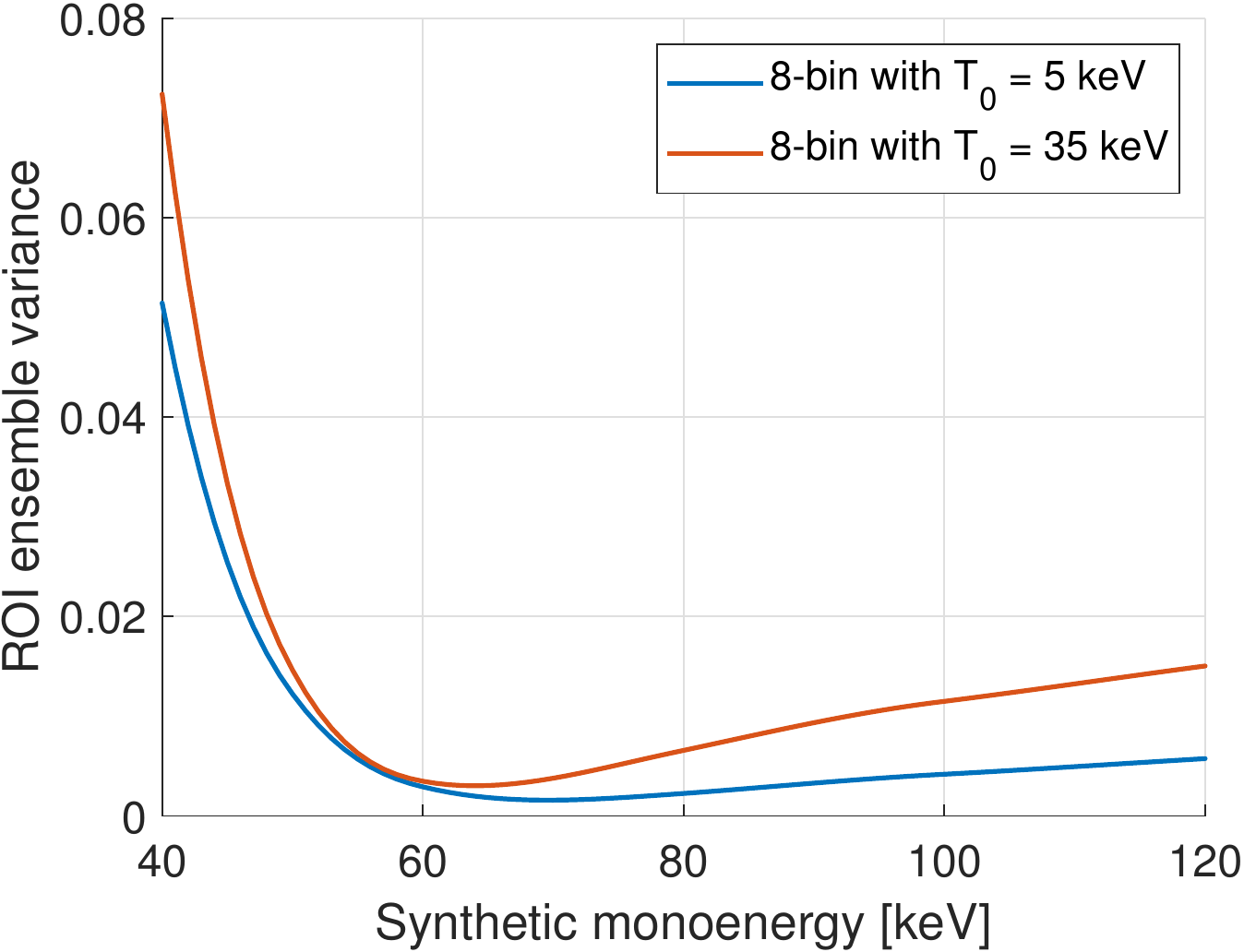}}
        \hfil
        \subfloat[]{\includegraphics[width=0.4\linewidth]{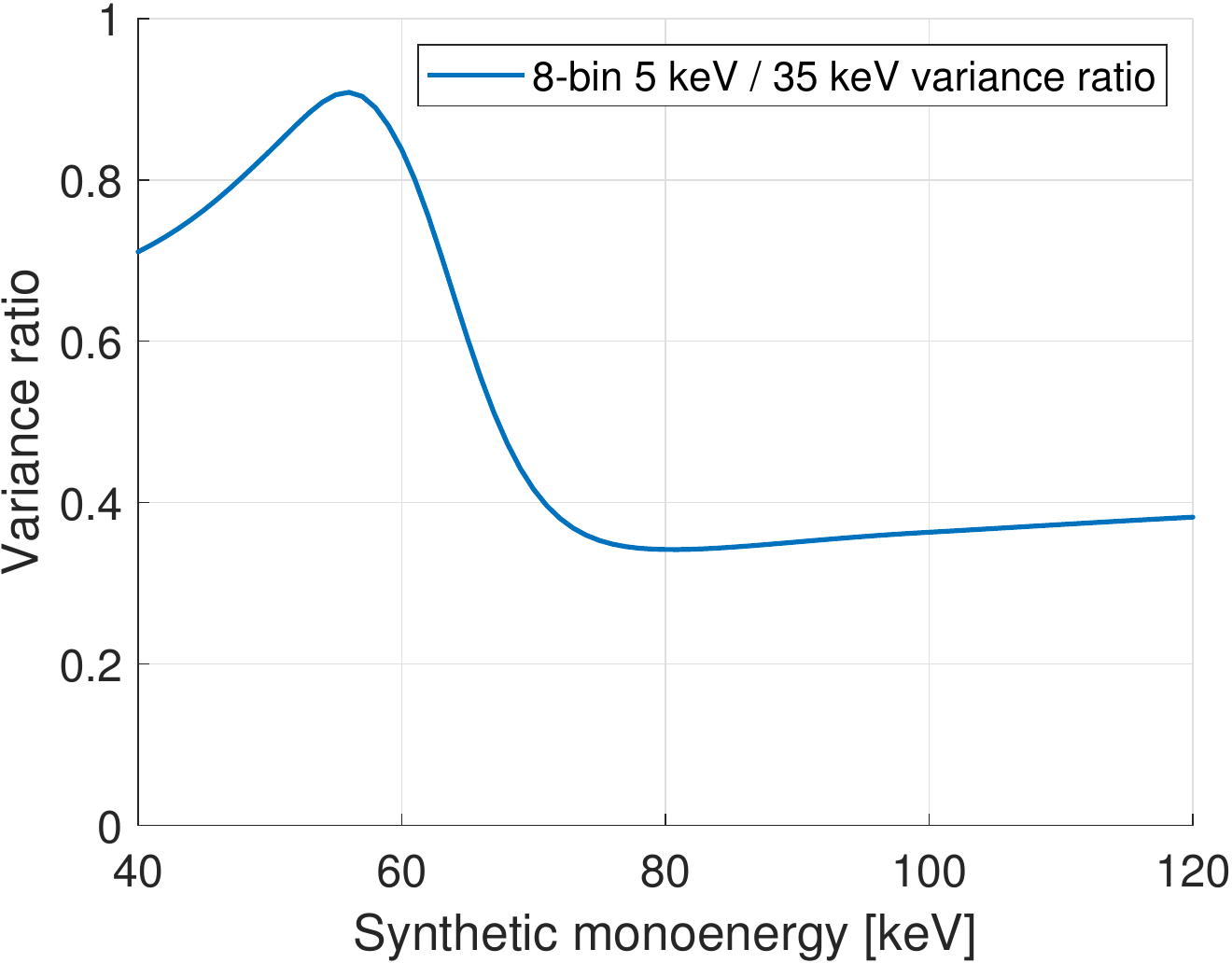}}          
        \end{minipage}
        \caption{(a) Noise variance in synthetic monoenergetic images for the simulated 8-bin \ds\ detector assuming a lowest threshold of 5~keV (solid, blue) and 35~keV (solid, red). (b) Noise variance ratio of the simulated 8-bin \ds\ detectors assuming lowest thresholds of 5~keV and 35~keV.}
        \label{fig:monoenergetic_variance}
  \end{figure*}

  The comparison of the Geometric-Poisson model with the Monte Carlo model for a pure photon\ch{-counting} detector is shown in Fig. \ref{fig:dqe_penalty_breakdown}. The projection domain metrics $\text{DQE}^{\,\text{\,density}}_{\,\text{projection}}$ and $\text{DQE}^{\,\text{\,spectral}}_{\,\text{projection}}$ are presented in Figs. \ref{fig:dqe} (a) and (b) and the image domain metrics $\text{DQE}^{\,\text{density}}_{\,\text{image}}$ and $\text{DQE}^{\,\text{spectral}}_{\,\text{image}}$ \ch{appear} in Figs. \ref{fig:dqe} (c) and (d), where they are compared with the projection domain metrics. Samples of reconstructed synthetic monoenergetic images at 40, 70 and 100~keV for the simulated 8-bin-\ds\ detector with lowest thresholds of 5~keV and 35~keV are \ch{shown} in Figs. \ref{fig:mono_images} (a)-(f). \ch{These figures}  illustrate the effect \ch{on image noise} of rejecting all multiple count \ch{events}.  We also present the ROI ensemble variance as a function of synthetic monoenergy for the simulated 8-bin \ds\ detector with lowest thresholds of 5~keV and 35~keV in Fig. \ref{fig:monoenergetic_variance} (a) and the corresponding variance ratio in Fig. \ref{fig:monoenergetic_variance} (b).

\section{Discussion}\label{sec:Discussion}

  In this work we investigated the effect of Compton scatter in \ds\ \ch{detectors} on zero-frequency DQE. Compton scatter is often assumed to be purely detrimental to the performance of \ds\ detectors, but as we have demonstrated, primary events from Compton interactions carry significant information for both density and spectral imaging tasks. This effect can be understood qualitatively from the simplified model of multiple counting developed in Section \ref{sec:geometric_poisson} and \ch{from} the results \ch{of} Fig. \ref{fig:dqe_penalty_breakdown} \ch{that  demonstrate the} trade-off between counting as many primary events as possible and avoiding extra noise variance caused by multiple counting of unique photons. \ch{At the particular} level of scatter in the simulated detector, the improved contrast \ch{imparted} by the primary Compton events outweighs the noise penalty from \ch{event multiplicity}. 

  As seen in Fig. \ref{fig:dqe}, similar conclusions hold for the simulated density imaging task: the benefit of counting primary Compton events \ch{also} outweighs the noise penalty from multiple \ch{count events} for lowest threshold energies down to 0~keV in both projection and image domains, as is shown in Fig. \ref{fig:dqe} (a) and (c). This indicates that setting the lowest threshold as low as possible is essential for density imaging performance of \ds\ detectors. Assuming a lowest threshold of 5~keV is consistent with an RMS noise level of 1.6~keV, which has been measured experimentally for \ds\ detectors \cite{xu2013energy,liu2014silicon}. For a lowest threshold of 5~keV, the density DQE was found to be 0.51 in both projection and image domains, with little \ch{or} no difference between assuming an unlimited number of energy bins \ch{versus limiting the number of energy bins to} 8.

  For the simulated spectral imaging task, the DQE plateaus below 10~keV, as \ch{illustrated} in Fig. \ref{fig:dqe} (b) and (d), indicating that photon interactions in silicon depositing less than 10~keV do not carry much spectral information. There is, however, a significant drop in spectral DQE between 10 and 30~keV, indicating that these interactions, which are almost all ($>99\%$) either Compton events or scatter, do carry important spectral information. For a lowest threshold of 5~keV, the spectral DQE was found to be 0.28 in the projection domain assuming an unlimited number of energy bins, and 0.26 assuming 8 energy bins, showing that there is a \ch{only a} slight reduction in spectral imaging performance from having a limited number of energy bins. 

  The importance of including Compton events can be seen visually in Fig. \ref{fig:mono_images}, where synthetic monoenergetic images from the simulated 8-bin \ds\ detector with a lowest threshold of 5 keV are compared with the corresponding images \ch{generated} using a threshold of 35~keV, which almost completely excludes Compton events. The noise variance as a function of monoenergetic energy for both detectors and the corresponding variance ratio are shown in Fig. \ref{fig:monoenergetic_variance} (a) and (b). \ch{It is apparent that the} degradation due to not including Compton events is greatest for higher monoenergies, and the smallest from 50--60~keV. This is consistent with the images in Fig. \ref{fig:mono_images}, where the 5~keV detector images exhibit 29\% lower variance at 40~keV, 58\% lower variance at 70~keV and 64\% lower variance at 100~keV compared to the detector using a 35~keV lowest threshold.  

  As can be seen in Fig. \ref{fig:dqe}, there is good agreement between the projection-domain and image-domain DQE for both density and spectral imaging. The largest relative discrepancy in the plotted energy range is 4.1\% for density DQE and 2.5\% for spectral DQE and the qualitative behaviour when varying the lowest threshold is similar, demonstrating that the projection-domain DQE provides a useful metric for predicting image quality in situations where including the reconstruction process in the analysis is impractical due to its mathematical complexity and computational cost. 

  The present work is focused on the effects of intra-detector Compton scatter and other sources of cross-talk on the zero-frequency DQE of the detector, which approximately reflects the performance for detection of large features. Compton scatter also gives rise to a point-spread function with long tails, which may cause shading effects near interfaces in the image. These tails can be corrected for through deconvolution \ch{so as not to cause}  artifacts in the image. Although the zero-frequency DQE studied here is not impacted by a linear deconvolution operation, further studies will be necessary to understand if the point-spread function and any subsequent deconvolution will have any substantial impact on the visual appearance of the image.  

  To fully assess the effect of Compton scatter on the detection performance, a more detailed analysis incorporating the modulation transfer function, the noise power spectrum of the reconstructed image, and a numerical observer able to mimic human observer performance \ch{are} needed. Although steps in this direction have been taken \cite{rajbhandary2020detective, persson2020detective}, the relationship between the image domain and projection domain performance for spectral imaging is still incompletely understood. An important topic for future investigations is there to extend the present study of image-domain and projection-domain DQE to nonzero spatial frequencies.

\section{Conclusion}\label{sec:Conclusion}

\ch{Owing to its low stopping power, silicon is not an obvious choice as a detector material for photons in the energy range used in human whole-body CT. However, an edge-on-irradiated geometry effectively compensates for the low linear attenuation coefficient of silicon by providing a long pathlength within the detector.}

\ch{The high Compton scatter fraction in low atomic number (Z) materials also appears to detract from the obviousness of using silicon in this application. However, as our simulations demonstrate, the vast majority of Compton scattered photons deposit energies below 30~keV, which is substantially separated from the range of useful photopeak energies. In contrast, the analogous spectral tailing due to K-escape fluorescence in high atomic number detectors (such as CdTe and CdZnTe) substantially overlaps with those energies that impart the highest value clinical information: photons in the range of \chd{$\approx$} 30--60~keV that are crucial for obtaining high soft tissue contrast, and contrast due to injected contrast agents such as iodine.}

\ch{The somewhat counterintuitive finding of this study, that the inclusion of Compton scatter events not only increase density DQE but also spectral DQE, further strengthens the case for using edge-on-irradiated low-Z detector materials for clinical CT.}

\section*{Acknowledgements}
This study received funding from the European Union’s Horizon 2020 research and innovation programme under the Marie Sklodowska-Curie grant agreement No. 795747, from the Swedish Research Council under grant No. 
2021-05103, from MedTechLabs and from GE Healthcare.
\noindent

\section*{Conflict of Interest}
Fredrik Grönberg discloses past financial interests in Prismatic Sensors AB and is currently employed by GE Healthcare. Mats U. Persson discloses past financial interests in Prismatic Sensors AB and research collaboration with GE Healthcare. Norbert J. Pelc is a consultant to GE Healthcare.
\noindent

\appendix
\section*{Appendix}

\section{Adding noise and binning}\label{sec:appendix_a}
The unbinned point-spread function and autocovariance function with electronic noise are obtained by  
\begin{equation}
  h^{\text{ub},\sigma}_{\boldsymbol{n}}(E,E') = \int_\mathbb{R} h^{\text{ub}}_{\boldsymbol{n}}(E,\E)g(E'-\E)d\E
\end{equation}
and
\begin{equation}
  K^{\text{ub},\sigma}_{\boldsymbol{n}}(E,E',E'') = \left\{\begin{aligned}&\int_\mathbb{R}\int_\mathbb{R} K^{\text{ub}}_{\boldsymbol{n}}(E,\E,\E')g(E'-\E)g(E''-\E')d\E d\E',&&\boldsymbol{n}\neq0 \\ &\int_\mathbb{R} K^{\text{ub}}_{\boldsymbol{n}}(E,\E,\E)g(E'-\E)d\E ,&&\boldsymbol{n}=0, E'=E''\\
  &0, &&\boldsymbol{n}=0, E'\neq E''\end{aligned}\right.
\end{equation}
For a given set of energy thresholds $T_1,\dots,T_M$ and the convention that $T_{M+1}=\infty$, the corresponding energy bin point-spread function and autocovariance function are obtained by
\begin{equation}
  h_{\boldsymbol{n},j}(E) = \int_{T_j}^{T_{j+1}} h^{\text{ub},\sigma}_{\boldsymbol{n}}(E,\E)d\E
\end{equation}
and
\begin{equation}
  K_{\boldsymbol{n},j,j'}(E) = \int_{T_j}^{T_{j+1}}\int_{T_{j'}}^{T_{j'+1}} K^{\text{ub},\sigma}_{\boldsymbol{n}}(E,\E,\E')d\E d\E'.
\end{equation}


\section*{References}

\bibliographystyle{medphy}

\end{document}